\documentclass{optica-article}
\usepackage{siunitx}
\usepackage[bottom]{footmisc}
\newcolumntype{?}[1]{!{\vrule width #1}}

\journal{opticajournal} 

\articletype{Research Article}

\usepackage{lineno}
\usepackage{enumitem}
\setlist[itemize]{noitemsep, topsep=0pt}

\begin{document}

\title{Learned, Uncertainty-driven Adaptive Acquisition for Photon-Efficient Scanning Microscopy}

\author{Cassandra Tong Ye\authormark{1}, Jiashu Han\authormark{2}, Kunzan Liu\authormark{1}, Anastasios Angelopoulos\authormark{3}, Linda Griffith\authormark{4}, 
Kristina Monakhova\authormark{1,5}, Sixian You\authormark{1,6}}

\address{\authormark{1}Research Laboratory of Electronics (RLE) in the Department of Electrical Engineering and Computer Science, Massachusetts Institute of Technology. Cambridge, MA USA 02139\\
\authormark{2}Fu Foundation School of Engineering and Applied Science, Columbia University in the City of New York. New York, NY USA 10027\\
\authormark{3}Department of Electrical Engineering and Computer Science, University of California, Berkeley. Berkeley, CA USA 94720\\
\authormark{4}Department of Biological Engineering, Massachusetts Institute of Technology. Cambridge, MA USA 02139
}

\email{\authormark{5}Kristina Monakhova: monakhova@cornell.edu\\\authormark{6}Sixian You: sixian@mit.edu } 

\begin{abstract*}
\noindent{}Scanning microscopy systems, such as confocal and multiphoton microscopy, are powerful imaging tools for probing deep into biological tissue. However, scanning systems have an inherent trade-off between acquisition time, field of view, phototoxicity, and image quality, often resulting in noisy measurements when fast, large field of view, and/or gentle imaging is needed. Deep learning could be used to denoise noisy microscopy measurements, but these algorithms can be prone to hallucination, which can be disastrous for medical and scientific applications. We propose a method to simultaneously denoise and predict pixel-wise uncertainty for scanning microscopy systems, improving algorithm trustworthiness and providing statistical guarantees for deep learning predictions. Furthermore, we propose to leverage this learned, pixel-wise uncertainty to drive an adaptive acquisition technique that rescans only the most uncertain regions of a sample, saving time and reducing the total light dose to the sample. We demonstrate our method on experimental confocal and multiphoton microscopy systems, showing that our uncertainty maps can pinpoint hallucinations in the deep learned predictions. Finally, with our adaptive acquisition technique, we demonstrate up to 16$\times$ reduction in acquisition time and total light dose while successfully recovering fine features in the sample and reducing hallucinations. We are the first to demonstrate distribution-free uncertainty quantification for a denoising task with real experimental data and the first to propose adaptive acquisition based on reconstruction uncertainty. 
\end{abstract*}

\section{Introduction}
Many popular microscopy modalities leverage scanning to probe deep into biological tissues; they focus light to a small region of a sample and collect light only from that region. 
By scanning light to different regions of the sample, they build up three-dimensional images of the sample often one point at a time. Scanning confocal microscopes have been widely adopted for medical and scientific applications due to their ability to recover three-dimensional structures by optical sectioning~\cite{wilson1990confocal}. Two-photon and multiphoton microscopy leverage non-linear excitation to probe even deeper into thick, scattering tissue. These systems are often used by neuroscientists to measure calcium dynamics in deep scattering mouse brains, as well as to characterize multicellular dynamics in immunology and cancer studies~\cite{attardo2015impermanence, lu2017video, prevedel2016fast, perrin2020frontiers}. In addition, label-free multiphoton microscopy enables minimally invasive imaging of biological structures in living and unlabeled biosystems, such as collagen fibers, immune cells, endothelial cells, and extracellular vesicles~\cite{you2021label}, through second harmonic generation (SHG)~\cite{fuentes2019second}, third harmonic generation (THG)~\cite{debarre2006imaging, barad1997nonlinear}, and two-photon and three-photon autofluorescence (2PAF, 3PAF)~\cite{zipfel2003live,liu2024deep}.
This has become an increasingly popular tool for tissue and cell microscopy in neuroscience, immunology, and cancer research~\cite{periasamy2020special, CHEN20111001,skala2007vivo, brown2001vivo, you2018intravital}. 

\begin{figure}
    \centering
    \includegraphics[width=\linewidth]{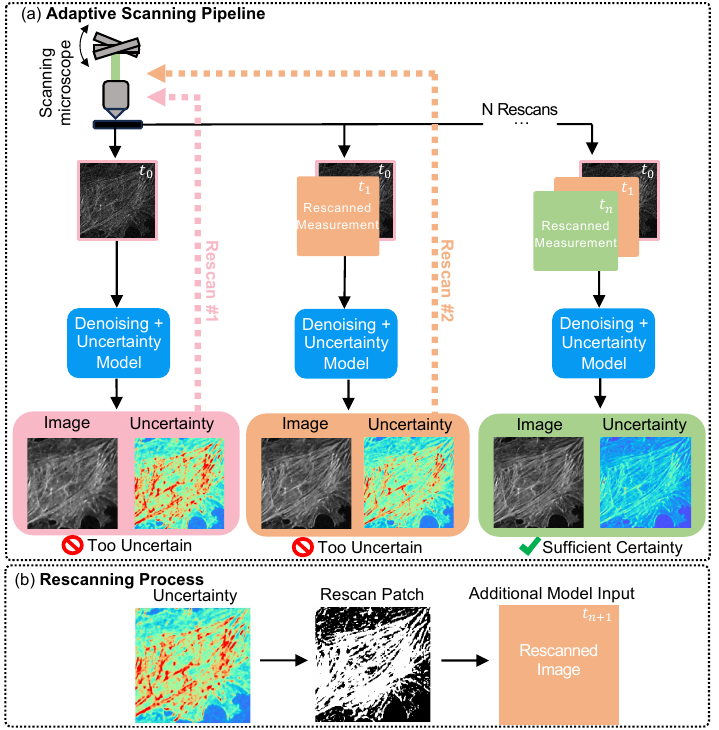}
    \captionsetup{width=1\linewidth}
    \caption{\textbf{Uncertainty-based Adaptive Imaging (a)}: A noisy measurement is acquired with a scanning microscopy system and passed into a deep learning model that predicts a denoised image and its associated pixel-wise uncertainty. Subsequently, the top N uncertain pixels are selected for a rescan, obtaining more measurements at only the uncertain regions. As more adaptive measurements are taken, the deep learning model predicts a denoised image with lower uncertainty. Scan duration and power are minimized, limiting sample damage while maintaining high confidence in the model prediction. \textbf{Rescanning Process (b)}: Given a pixel-wise uncertainty prediction, regions with high uncertainty can be selected for rescanning. Only this patch of pixels will be rescanned in the sample, and this patch, superimposed with the original, becomes an additional channel that is fed into the model.}
    \label{fig:Overview}
\end{figure}

To provide a minimally perturbative window into the tissue architecture and cell dynamics of intact biosystems, the next advances in scanning microscopy require deeper, faster, and gentler imaging of thick and living samples~\cite{volpe2023roadmap}. For scanning microscopy systems, there is an inherent trade-off between acquisition time, field of view, phototoxicity, and image quality, often resulting in noisy measurements when fast, large field of view, deep, and/or gentle imaging is needed. Noisy images can be challenging to interpret, and fine structures within the images can be obscured by noise. 

Deep learning-based methods have shown exciting results for denoising extremely noisy images in microscopy~\cite{weigert2018content, platisa2023high, von2021democratising}. Specifically for scanning microscopy, deep learning methods have demonstrated faster scan times and improved image quality~\cite{hsu2022three, zhang2019poisson, lee2020mu, laine2021imaging, manifold2019denoising}. Despite the success of deep learning, there is still a lot we do not understand, and the robustness of deep learning-based methods has been called into question~\cite{antun2020instabilities}. Specifically, deep learning methods have been shown to hallucinate~\cite{ji2023survey, helou2020stochastic, bhadra2021hallucinations}, i.e., produce realistic-looking artifacts that are not truly present in the sample, and suffer from instabilities, i.e., sensitivity to small perturbations in the measurement~\cite{antun2020instabilities}. Hallucination may be acceptable or even desirable for photography or image generation but is unacceptable for biomedical and scientific imaging, where small image features may lead to important diagnoses or discoveries. To be adopted for scientific and medical imaging, deep learning methods need to be robust and trustworthy.     

Uncertainty quantification techniques can help catch model hallucinations and improve the robustness of deep learning methods. Typically, a deep learning model will be trained using image pairs (e.g., noisy and noiseless), and then, at test time, given a measurement, the model will produce an estimate of the image. In addition to the image estimate, uncertainty quantification methods aim to produce an uncertainty map for each pixel in the image (Fig.~\ref{fig:Overview}). Typically, this has been accomplished with Bayesian neural networks with techniques like dropout~\cite{gal2016dropout, abdar2021review, Xue_2019}. More recently, conformal risk control paired with pixel-wise quantile regression has been proposed for distribution-free uncertainty quantification~\cite{angelopoulos2022image}.
This method is much faster than Bayesian approaches, requiring only one forward pass at test time, and produces confidence sets backed by rigorous statistical guarantees. 
\begin{figure}
    \centering
    \includegraphics[width=\linewidth]{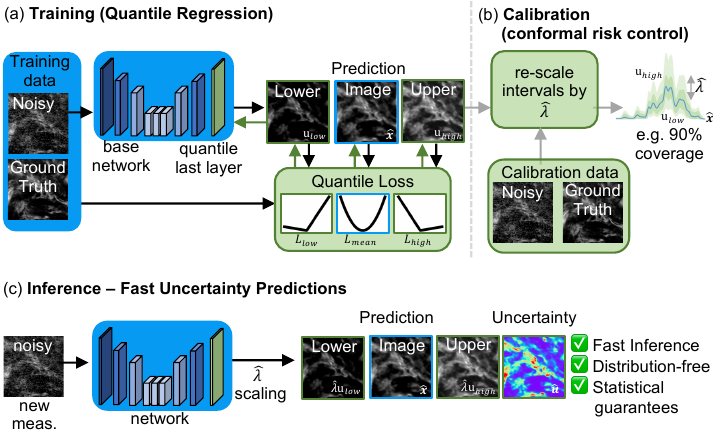}
    \captionsetup{width=1\linewidth} 
    \caption{\textbf{Learned Uncertainty Quantification}: To leverage distribution-free uncertainty quantification with a deep network, three modifications are needed: a last layer that predicts a lower and upper bound, a modified loss, and a post-training calibration step. Standard deep denoising steps are highlighted in blue, while our modifications are highlighted in green. (a) During training, the modified network returns three channels: the lower uncertainty quantile, the denoised prediction, and the upper uncertainty quantile. A quantile loss function defines the loss for the upper and lower quantiles, encouraging underestimates and overestimates. (b) After training, a calibration step is needed to adjust the upper and lower bounds and provide statistical guarantees for this predicted interval. (c) At test time, a single forward pass determines the uncertainty at each step, which is the difference between the upper and lower quantile predictions.}
    \label{fig:Architecture}
\end{figure}

In our work, we propose a multi-frame uncertainty quantification technique for scanning microscopy denoising and an adaptive microscopy imaging pipeline informed by uncertainty quantification (Fig.~\ref{fig:Overview}). To quantify uncertainty, we train a network to predict a lower and upper bound for each image using quantile regression~\cite{koenker1978regression}. Then, we use conformal risk control,  a post-training calibration step, to properly scale our learned uncertainty intervals so that they contain the correct fraction of ground truth pixels based on a held-out calibration dataset, (Fig.~\ref{fig:Architecture})~\cite{angelopoulos2022conformal, angelopoulos2022image}. With these modifications, the denoiser will predict not only a denoised image but also a confidence interval for each pixel with a statistical guarantee that a percentage of the true values will fall within this interval. From this confidence interval, we can identify regions with the highest uncertainty, which can be caused by uncertainty in the model or signal. Having a confidence interval with statistical guarantees that the true signal intensity lies within these bounds is a significant advance towards making deep-learning-based denoising methods trustworthy enough for scientific and medical imaging.

Taking it one step further, we propose to leverage the learned uncertainty to drive adaptive acquisition: we capture more measurements of our sample \textit{only} at the most uncertain regions rather than rescanning the whole sample. Thus, we first capture a single noisy scan (fast and minimal sample damage due to low power or dwell time), followed by denoising, and then repeated adaptive scanning of the most uncertain regions. This way, we can leverage signal across multiple frames to improve denoising performance and reduce uncertainty while minimizing scan duration and damage to the sample. Our proposed adaptive acquisition is both sample and algorithm-informed; i.e., we acquire more measurements in locations where our denoising algorithm is struggling in order to resolve finer features, reduce hallucinations, and improve overall confidence in the image prediction. 


Our main contributions include:

\begin{itemize}
    \item Adapting distribution-free uncertainty quantification based on conformal risk control to multi-image denoising
    \item Evaluating the performance of multi-image denoising with uncertainty quantification on real, experimental scanning microscopy data (confocal, two-photon, and multiphoton)
    \item Proposing and evaluating an adaptive acquisition technique for scanning microscopy based on conformal risk control 
    \item Demonstrating a potential 1-16$\times$ improvement in scanning speed and total light dose using adaptive acquisition in a software-only demonstration using real data.
\end{itemize}


\section{Related Work}

\subsection{Denoising}
A variety of denoising techniques have been proposed throughout the years. Classic denoising methods rely on hand-picked priors and heuristic design, such as sparsity~\cite{portilla2003image}, smoothness, or patch similarity~\cite{dabov2007image}. These methods have been surpassed by data-driven methods based on deep learning~\cite{zhang2017beyond}; however, these methods trade accuracy for stability~\cite{gottschling2020troublesome} as well as perception for distortion~\cite{blau2018perception}. Recently, the concept of diversity denoising, which attempts to obtain a diverse set of different plausible restorations from a single corrupted measurement, has been proposed to deal with the problem that multiple solutions exist to ill-posed inverse problems~\cite{prakash2020fully2, prakash2021interpretable}. This method is promising and helps gauge plausible features within a prediction, but it is computationally intensive, requiring many forward passes, while ours directly predicts uncertainty.

Specifically for low-light measurements with poor SNR, leveraging multiple frames rather than a single frame during denoising can improve performance since each measurement will have correlated signal but uncorrelated noise. Burst denoising, in which multiple noisy frames are used to estimate a single noiseless image, is used for nighttime photography on smartphones~\cite{liba2019handheld}. Similarly, leveraging multiple frames for video denoising has demonstrated impressive results in extreme low light conditions~\cite{monakhova2022dancing}. Unsupervised and self-supervised burst denoising has shown remarkable results for calcium imaging by adapting the popular Noise2Self~\cite{batson2019noise2self} and Noise2Void~\cite{krull2019noise2void} frameworks to handle multiple frames~\cite{lecoq2021removing, li2021reinforcing}. We take inspiration from burst denoising methods to develop an adaptive scanning approach that adaptively acquires new measurements to improve our denoising performance and reduce uncertainty.




\subsection{Uncertainty quantification methods}
Uncertainty quantification for deep learning is an active area of interest, and is particularly critical for leveraging machine learning models in physics, science, and medical imaging~\cite{gal2022bayesian, abdar2021review, zhang2021blindnet}. The most popular methods are based on Bayesian statistics, including methods such as Bayesian networks and Monte Carlo dropout~\cite{gal2016dropout,kendall2017uncertainties, neal2012bayesian, shang2023approximating}, or are ensemble-based~\cite{NIPS2017_9ef2ed4b}, i.e. training multiple networks as a proxy for uncertainty. However, these methods have two main issues: they cannot provide strong statistical guarantees (a.k.a. they cannot guarantee how reliably they evaluate the true variability of the underlying data distribution, often requiring accurate distribution priors)~\cite{hirsh2022sparsifying, freeman2015improving, xue2019reliable}, and are computationally intensive, requiring multiple passes through a network at test time. 

More recently, distribution-free uncertainty quantification based on conformal risk control have been proposed as an alternative to Bayesian methods for uncertainty quantification in deep networks~\cite{angelopoulos2022image, angelopoulos2021gentle, teneggi2023trust, cheung2024metric}. This approach predicts pixel-wise uncertainty intervals for each image with a user-specified confidence probability. This method is agnostic to model or data distributions, provides formal mathematical guarantees, and is computationally inexpensive, requiring only a single forward pass. We leverage distribution-free uncertainty quantification for scanning microscopy denoising, evaluating its performance on a challenging experimental denoising problem for the first time.


\subsection{Next best view and active computer vision}
The research areas of `next best views'~\cite{connolly1985determination} and `active vision'~\cite{aloimonos1988active} in robotics and computer vision both deal with how to plan future acquisitions given some measurements and a guiding high-level task, such as 3D reconstruction or localization. There have been a number of methods developed to determine and plan the next best view based on factors such as occlusion~\cite{maver1993occlusions}, uncertainty in 3D reconstructions~\cite{dunn2009developing}, and camera and lighting conditions to deal with scattering media~\cite{sheinin2016next}. Many of these methods seek to improve information gain or reduce uncertainty in reconstructions by actively determining which measurements to take. However, many assume distributions and explicit models for uncertainty. Our work is related to this research area, since it involves path planning (the scanning trajectory of the microscope) and aims to choose the next best measurement based on reconstruction uncertainty. However, we do not make any assumptions about the noise or measurement distribution and leverage conformal risk control to predict and update the uncertainty bounds. Conformal risk control is actively being explored for making safe autonomous decisions from imperfect predictions~\cite{lekeufack2024conformal}, and we believe that our method could have parallels and insights for the next best view and active vision communities. 



\subsection{Adaptive acquisition in microscopy}
A number of adaptive acquisition techniques have been proposed to reduce overall light dose, reduce scanning time, or enhance weak signals in microscopy~\cite{hoebe2007controlled, abouakil2021adaptive, dreier2019smart, chu2007enhanced}. Many of these methods rely on detecting the sample or regions of interest, either by a coarse prescan~\cite{abouakil2021adaptive, li2020adaptive} or real-time specimen tracking~\cite{royer2016adaptive, mcdole2018toto}. Several event-driven approaches have been proposed to adapt acquisition from a slow to a faster imaging rate after detecting an event either heuristically~\cite{alvelid2022event} or with a neural network trained to detect precursors to events~\cite{mahecic2022event}. In addition, machine learning techniques have been utilized to optimize the excitation power for a 3D sample in MPM~\cite{pinkard2021learned}. While previous methods decouple the adaptive acquisition from any reconstruction algorithms (or lack thereof), our method is distinct in that we tightly couple the adaptive acquisition and the reconstruction algorithm. Our adaptive acquisition is driven by uncertainty in our denoising algorithm, with each additional measurement improving the algorithm's performance. This allows us to push the trade-offs of light dose, speed, and imaging quality by adaptively imaging at low signal levels while algorithmically recovering a high-fidelity signal. 







\section{Methods}
Our goal is to predict a noiseless image and its uncertainty given one or several measurements that are corrupted by noise. Assuming our sample is non-moving, our measurements can be described as: 

\begin{equation}
    \mathbf y_t = n_t (\mathbf x), 
\end{equation}

\noindent where each measurement $\mathbf y_t$ contains the signal, $\mathbf x$, which is measured through some noise function $n_t(\cdot)$. Our denoiser, $F_{\theta}$, predicts both the denoised image, $\hat{\mathbf x}_T$, and the uncertainty, $\hat{\mathbf u}_T$, at each pixel:

\begin{equation}
    \hat{\mathbf x}_T, \hat{\mathbf u}_T = F_{\theta}(\mathbf y_0, ..., \mathbf y_T).
\end{equation}

For single-image denoising, the denoiser uses a single measurement, $\mathbf y_0$. For multi-image denoising, the denoiser uses $T$ measurements $\mathbf y_0, ..., \mathbf y_T$. Using multiple measurements improves the prediction and reduces uncertainty at each pixel. We can use our denoiser as is to obtain a denoised image and uncertainty estimate, or we can use the uncertainty to drive our scanning process for adaptive microscopy. 

For adaptive microscopy, we take a preliminary scan of the sample, then feed this measurement into the denoiser to obtain the initial image and uncertainty estimates, $\hat{\mathbf x}_0, \hat{\mathbf u}_0 = F_{\theta}(\mathbf y_0)$. The uncertainty estimate at each pixel is used to determine which locations in the sample to rescan next:

\begin{equation}
\label{eq:adaptive}
    (\mathbf{r}_{t+1}, \mathbf{c}_{t+1}) = A(\hat{\mathbf u}_t),
\end{equation}

\noindent where $r$ and $c$ represent the rows and columns of the locations in the sample. Thus, the uncertainty estimate at time $t$ determines the pixels to scan at time $t+1$. The most uncertain areas of the sample are rescanned to obtain the next measurement, $\mathbf y_{t+1}(\mathbf{r}_{t+1}, \mathbf{c}_{t+1}$). The function $A(\cdot)$ defines our re-sampling strategy based on the pixel-wise uncertainty. The simplest update strategy is to rescan pixels above a certain uncertainty threshold. A more sophisticated update strategy would consider the dynamics and physical constraints of the scanning hardware, for example, rescanning rows at a time rather than random-access pixels. 

With this adaptive microscopy approach, we rescan only the most uncertain positions in our sample, saving time and reducing photodamage to our sample. With each subsequent scan, the fidelity of our denoised image estimate improves, and our uncertainty decreases, meaning we need to rescan fewer positions each time. In the following sections, we detail how to add uncertainty quantification into a denoising network and how we design the update rules for adaptive microscopy. 

\subsection{Learning uncertainty through quantile regression}
Our adaptive microscopy method is based on rescanning the most uncertain regions of the sample, but how do we know where the most uncertain regions are? In this work, we use pixel-wise quantile regression as a metric for uncertainty~\cite{koenker1978regression}. First, we need a network that can predict the lower bound and the upper bound, in addition to predicting each denoised pixel. We can do this by modifying a deep network to return three channels instead of one (Fig.~\ref{fig:Architecture}a) or by doing three forward passes through a network with an additional control input to specify the desired quantile (Fig. S1). This will give us three separate images for each set of measurements. 

\begin{figure}
    \centering
    \includegraphics[width=\linewidth]{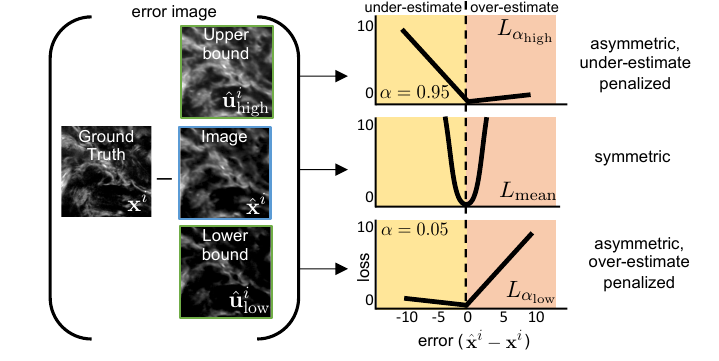}
    \captionsetup{width=1\linewidth} 
    \caption{\textbf{Quantile regression}. An asymmetric loss function is used to encourage the deep network to predict the upper and lower-bound images. As opposed to a symmetric loss like MSE, which penalizes estimates above and below the mean value equally, quantile loss is asymmetric. When $\alpha$ is small (bottom), predictions higher than the mean (over-estimates) are more heavily penalized than predictions lower than the mean (under-estimates), encouraging the network to predict an image that is below the mean (lower bound image). When $\alpha$ is large (top), under-estimates are more heavily penalized than under-estimates, encouraging the network to predict an image that is above the mean (upper bound image). }
    \label{fig:quantile_loss}
\end{figure}

If we want our lower and upper-bound images to form a confidence interval with 90\% coverage, i.e., a 90\% probability that the confidence interval will include the true value, then the lower and upper-bound images should predict the 5\% quantile and 95\% quantile images, respectively. Common image regression losses, such as mean squared error (MSE) or the L1 loss are symmetric, Fig~\ref{fig:quantile_loss}(middle), which means that image predictions that are larger or smaller than the ground truth image are equally penalized, encouraging a network to predict the mean image. To encourage the network to predict the lower and upper-bound images, we use a quantile loss, which is asymmetric. The quantile loss is defined as:

\begin{equation}
    L_{q}(\mathbf x^i, \hat{\mathbf x}^i) = \begin{cases}
q \cdot \lvert \mathbf x^i - \hat{\mathbf x}^i \rvert & \text{if } \mathbf x^i -\hat{\mathbf x}^i \geq  0 \\
(1-q) \cdot \lvert \mathbf x^i - \hat{\mathbf x}^i \rvert & \text{otherwise,}
\end{cases}
\end{equation}
\noindent where $q$ is the desired quantile, ranging from 0 to 1, $\hat{\mathbf x}^i$ is the predicted image, and $\mathbf x^i$ is the ground truth image. When $q > 0.5$ or $q < 0.5$, the loss function becomes asymmetric and has a higher penalty for estimates below or above the ground truth, respectively, Fig.~\ref{fig:quantile_loss}. Thus, a predicted image that is smaller than the ground truth (under-estimate) will have a larger penalty than a predicted image that is larger than the ground truth (over-estimate) when $q > 0.5$ (Fig.~\ref{fig:quantile_loss} (top)). This will encourage the network to predict an image that's larger than the mean (upper-bound image). Conversely, when $q < 0.5$, an over-estimate will have a larger penalty than an under-estimate, which encourages the network to predict an image that is smaller than the mean (lower-bound image) (Fig.~\ref{fig:quantile_loss} (bottom)). 

Thus, quantile loss can be used to encourage the network to predict the upper-bound and lower-bound images. For our training model, we used $q_{low} = 0.05$ and $q_{high} = 0.95$ to obtain 90\% coverage for our confidence interval. These values can be adjusted to obtain higher or lower coverage. 

While quantile loss is used for the upper and lower bound images, any standard image loss can be used for image prediction (e.g., MSE, L1, LPIPS). Thus, the three predicted images share network weights but use a different loss function: mean squared error (MSE) for the image prediction and quantile loss for the lower and upper bounds.  Training can be summarized as solving the following optimization problem: 

\begin{equation}
\min_{\theta} \sum_i L_{q_{\text{low}}}(\mathbf x^i,\hat{\mathbf u}_{\text{low}}^i) +  L_{\text{mean}}(\mathbf x^i,\hat{\mathbf x}^i) +L_{q_{\text{high}}}(\mathbf x^i,\hat{\mathbf u}_{\text{high}}^i) .
\end{equation}

\noindent Where $\hat{\mathbf x}^i = F_{\theta}(\mathbf y_0^i, ..., \mathbf y_T^i)$ is the predicted denoised image and  $L_{\text{mean}}(\mathbf x^i,\hat{\mathbf x}^i) =  \| \mathbf x^i - \hat{\mathbf x}^i \|_2^2$ is the loss on the image prediction channel.  Here, ($\mathbf x^i, \mathbf y_0^i, ..., \mathbf y_T^i$) corresponds to image pairs within our training dataset consisting of one ground truth image for each set of noisy measurements, with a total of $N$ pairs. After training, our network, $\mathbf F_{\theta}$, produces a lower bound, image prediction, and upper bound. 


\subsection{Calibrating Uncertainty through Conformal Risk Control}
\label{sec:riskcontrol}
Now, we have an uncertainty interval at each pixel, but can we trust it? This is where conformal risk control comes in~\cite{angelopoulos2022image, angelopoulos2021gentle}. With conformal risk control, a held-out calibration dataset is used to assess and calibrate the uncertainty intervals, which are statistically guaranteed to cover the true labels with a user-specified coverage probability. This calibration step is distribution-free, meaning it does not rely on any assumptions about the distribution of our dataset or model. During calibration, the lower and upper bounds, $(\hat{\mathbf u}_{\text{low}}, \hat{\mathbf u}_{\text{high}})$,  are scaled by $\hat{\lambda}$, a scalar constant, until they contain the correct fraction of ground truth pixels based on the calibration dataset. The final pixel-wise uncertainty is:

\begin{equation}
    \hat{\mathbf{u}} = \hat{\lambda} ( \hat{\mathbf u}_{\text{high}} - \hat{\mathbf u}_{\text{low}} ).
\end{equation}

\noindent  To determine $\hat{\lambda}$, we select a maximum allowable percentage of incorrect values, $\alpha$, where $\alpha$ ranges between 0 and 1 (e.g., 0.1 means our set will contain 90\% of the true values). Given a user-specified $\alpha$, we can use the following algorithm to choose $\hat{\lambda}$:
\begin{equation}
    \hat{\lambda} = \inf \left\{ \lambda : \hat{R}(\lambda) \leq \alpha - \frac{1 - \alpha}{N} \right\},
\end{equation}

\noindent where $N$ is the size of the dataset, and $\hat{R}(\lambda)$ is the risk, which is defined as:

\begin{equation}
    \hat{R}(\lambda)=\frac{1}{N}\sum_{i=1}^{N}\frac{\text{\# of pixels miscovered in image}\: i \text{ when using } 
\lambda}{\text{total \# pixels in image} \: i}.
\end{equation}
The risk function gauges the predictive risk associated with the chosen $\lambda$ by calculating the number of pixels that are not within the confidence interval given ${\lambda}$, commonly called `miscovered pixels' since they are not covered by the confidence set. The risk is averaged across the entire calibration dataset.

For example, when $\alpha = 0.1$ and the dataset is large, $\hat{\lambda} = \inf \left\{ \lambda : \hat{R}(\lambda), \leq 0.1 \right\} $, which means that the scalar value $\hat{\lambda}$ will increase/decrease the bounds until at least 90\% of the true pixels in the calibration dataset fall within the predicted confidence range. With the same alpha and a smaller dataset, say N=20, $\hat{\lambda} = \inf \left\{ \lambda : \hat{R}(\lambda), \leq 0.055 \right\} $, so $\hat{\lambda}$ will need to be larger to ensure that at least 94.5\% of the pixels fall within the confidence range. 

This choice of $\hat{\lambda}$ statistically guarantees that the confidence interval will cover $1-\alpha$ fraction of future pixels in expectation, given that the calibration set and new test data points are exchangeable. The proof and analysis for this can be found in~\cite{angelopoulos2022image}. 


\subsection{Uncertainty-driven Adaptive acquisition}
Pixel-wise uncertainty is used to drive our adaptive microscopy scans. Given an initial measurement at time $t$, our model predicts the denoised image and its associated uncertainty, $\mathbf{u}_t$. To choose which coordinates to rescan at time $t+1$, we select coordinates within the image that have an uncertainty above the threshold $u_{\text{thresh}}$.
Towards this end, let $\mathcal{I}_{t+1} = \{ (r, c) : \mathbf{u}_t(r, c) \geq u_{\text{thresh}} \}$ be the set of all coordinates whose uncertainty is larger than the threshold.
The set $\mathcal{I}_t$ specifies the coordinate vectors fully, and they can be rewritten as
\begin{equation}
    \begin{aligned}
        &\mathbf{r}_{t+1} = \big(\mathcal{I}_{t+1}(0)(0), \ldots, \mathcal{I}_{t+1}(n_{t+1})(0)\big) \\
        &\mathbf{c}_{t+1} = \big(\mathcal{I}_{t+1}(0)(1), \ldots, \mathcal{I}_{t+1}(n_{t+1})(1)\big),
    \end{aligned}
\end{equation}
where $n_{t+1}$ represents the cardinality of $\mathcal{I}_{t+1}$. 

\noindent Thus, we only rescan the most uncertain pixels within the sample (Fig.~\ref{fig:Overview}b), with the uncertainty at time $t$ driving the adaptive scan at time $t+1$. In each subsequent pass, the model takes in the original noisy measurement and superimposed rescans to perform multi-image denoising. With each pass, our model has more data at the uncertain pixels and can improve its image prediction and confidence. This iterative process increases the number of observations at the most uncertain positions within the sample until the model's prediction is within an acceptable uncertainty level (Fig.~\ref{fig:Overview}a). In practice, this process is limited by the fixed, finite input size of the denoiser. In our case, we choose the network input size to be twenty and are therefore limited to nineteen total adaptive rescans. Note that the statistical guarantees from conformal risk control apply only to the first round of scanning — the guarantee is formally lost after multiple rounds of interval construction due to the distribution shift from adaptive scanning, but we find the practical effect of this to be limited.

\section{Implementation details}

We tested our uncertainty quantification and adaptive acquisition technique on an existing experimental confocal and two-photon dataset~\cite{zhang2019poisson}, as well as a custom experimental multiphoton autofluorescence data of fixed mouse whisker pad samples. Each dataset includes multiple noisy measurements and corresponding ground truth images that are obtained from averaging the noisy measurements. Both datasets contain only static, non-moving samples without any noticeable photobleaching. For adaptive acquisition, we synthetically rescan the images in software without any in-the-loop hardware rescanning.  



\subsection{Datasets}
For the experimental confocal and two-photon datasets, we used images from the Fluorescence Microscopy Denoising (FMD) dataset~\cite{zhang2019poisson}, which consists of 12,000 real fluorescence microscopy images of biological samples obtained with commercial confocal, two-photon, and wide-field microscopes. This dataset has Poisson-Gaussian noise and ground truth images are obtained by averaging 50 noisy measurements. From this dataset, we used 119 samples for training, 34 for calibration, and 7 for testing.

For our custom multiphoton dataset, we collected measurements from a custom-built inverted scanning microscope~\cite{liu2024deep}. The microscope utilizes a custom-built fiber source at 1100 nm~\cite{liu2024deep}, which excites multiphoton autofluorescence signals that are collected by individual photomultiplier tubes. We captured measurements of a fixed mouse whisker pad tissue at 64 sites within the sample, each time capturing 20 noisy measurements, which are averaged to obtain a reference ground truth image. Our final dataset contains multiphoton autofluorescence images from 32 sites for testing and 32 sites for calibration. See Suppl. Section 5 for more details. 


\begin{figure}
    \centering
    \includegraphics[width=\linewidth]{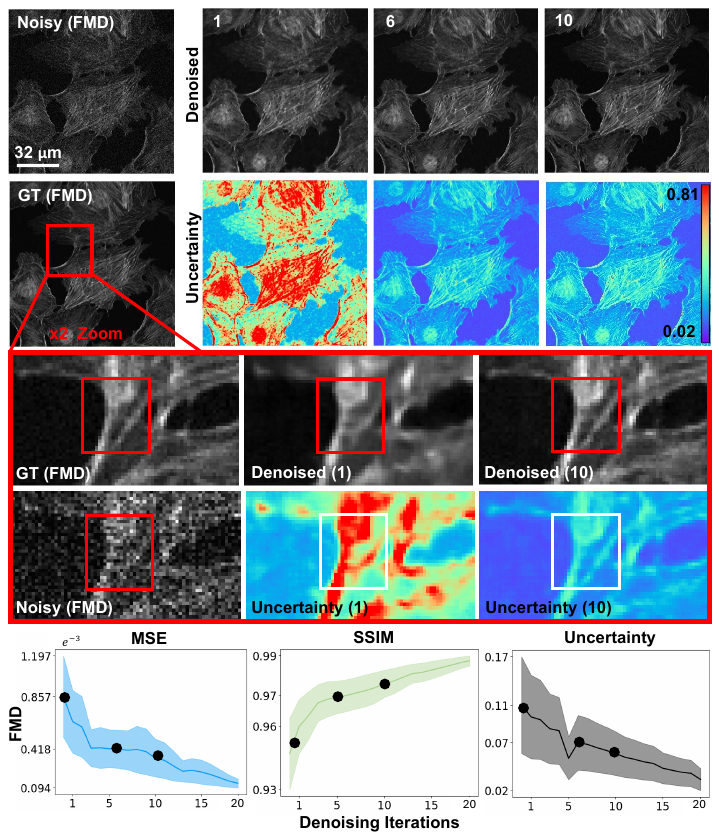}
    \captionsetup{width=1\linewidth}
    \caption{\textbf{Multi-image FMD Denoising Results}. We compare single and multi-image denoising on a representative two-photon sample from the FMD dataset. As the number of measurements increases from 1 to 6 and 10, sharper features emerge in the denoised images (top row) and the predicted uncertainty (2nd from top row) decreases. Quantitatively, the MSE and SSIM improve over iterations, and the average uncertainty decreases (bottom row). Zooming in on a small region within the sample (inset), we can see that single-image denoising produces a hallucination (red box): a horizontal feature that is not present in the ground truth image. This hallucination has high uncertainty. After multi-image denoising, the hallucination goes away as the predicted image converges to the ground truth feature, and concurrently the uncertainty in this region decreases.}
    \label{fig:FMD_denoising}
\end{figure}

\subsection{Networks, Training, and Calibration}
We used two different networks for denoising: NAFNet~\cite{chen2022simple} with a modified last layer to output 3 channels and a U-Net with a control input to select the quantile. Both networks produced similar results and hallucinations. For simplicity, we detail the NAFNet implementation and results in the main text, and the U-Net implementation and comparisons in the Suppl. We used the base NAFNet~\cite{chen2022simple} model with 20 inputs and modified the last layer of our network as detailed in Angelopoulos 2022~\cite{angelopoulos2022image} to produce the upper and lower quantile images (Fig.~\ref{fig:Architecture}). To simulate adaptively acquiring different images of the same sample, our training inputs were tensors of size 20 x 512 x 512, where each 512 x 512 tensor was originally repeated 20 times. As the iterations increase, different noisy images were substituted in and this occurs until all twenty inputs are different images of the same sample. 

The model was trained for 100 epochs on the FMD dataset using an Adam optimizer with an initial learning rate of 1e-3 and a cosine annealing learning rate schedule. We used the same network for both the FMD and multiphoton datasets. While our model performs well on FMD images, it also generalizes to multiphoton microscopy images. This occurs without the need for prior retraining or fine-tuning.

\subsection{Conformal Calibration}
After training, we calibrated the uncertainty interval using conformal risk control and our calibration datasets, as explained in Sec.~\ref{sec:riskcontrol}. From our acquired images, we reserve a few samples that are not used for training or testing, but specifically for the purposes of calibrating our uncertainty interval. We calibrate the uncertainty interval with a held-out calibration dataset because this provides us with a formal guarantee of our uncertainty interval coverage. We set $\alpha=0.1$ to select a 90\% confidence interval, finding $\hat{\lambda}$ respectively for each MPM channel and rescan iteration. To highlight darker regions and features within our images, we display gamma corrected resulted for select images (measurements, denoised, ground truth), excluding uncertainty images (images with a red and blue colormap)

\section{Results and Evaluation}
We evaluate our denoising and uncertainty prediction on an experimental dataset of confocal and two-photon measurements (FMD dataset) as well as a custom multiphoton dataset (MPM dataset). First, we evaluate our denoiser's performance as a function of the number of measurements. We show that our uncertainty prediction can help pinpoint a model hallucination that disappears once more measurements are utilized. Finally, we evaluate our uncertainty-driven adaptive acquisition technique, testing the performance against single and multi-image acquisition and quantifying the time and light-dose benefits. 

\subsection{Multi-image denoising and pinpointing hallucinations}
\begin{figure}
    \centering
    \includegraphics[width=\linewidth]{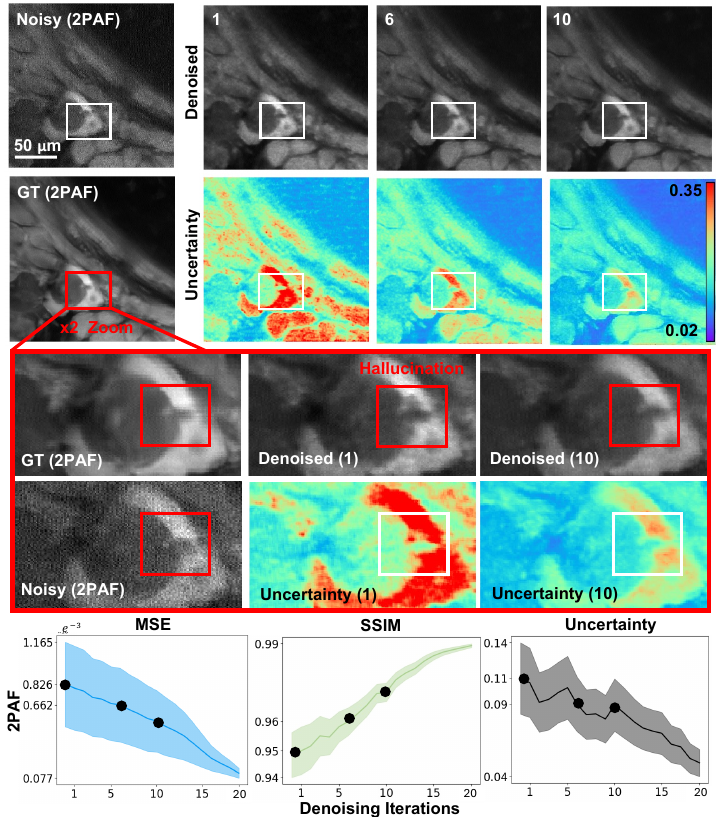}
    \captionsetup{width=1\linewidth}
    \caption{\textbf{Multi-image MPM Denoising Results}. Next, we compare single and multi-image denoising on a 2PAF sample from our custom MPM dataset, using a network that has never seen MPM data. As before, multi-image denoising results in sharper features and a lower uncertainty (top two rows), which can be quantified by the MSE, SSIM, and average uncertainty (bottom). Single-image denoising results in a hallucination (red box) which is not present in the ground truth. }
    \label{fig:MPM_denoising}
\end{figure}

First, we test our denoiser's performance as a function of the number of noisy measurements. To do this, we vary the input to the network, feeding in 1-20 unique, non-moving, noisy measurements. Each measurement has a different instance of noise, so intuitively, with more measurements, the denoising prediction should improve. We show the predicted denoised images and uncertainty maps for samples from the FMD dataset and MPM datasets in Figs.~\ref{fig:FMD_denoising}-\ref{fig:MPM_denoising}. Here, we can see that as the number of measurements increases, the network's predicted uncertainty decreases, and concurrently, the denoised predictions become sharper and closer to the ground truth. 

Single-image denoising produces a compelling prediction of the true signal, but upon closer inspection, it is evident that there are multiple areas of hallucination, i.e., realistic-looking features that are not present in the ground truth image. Examples of these features are highlighted in Figs.~\ref{fig:FMD_denoising}-\ref{fig:MPM_denoising} and  Figs. S1-S2. In Fig.~\ref{fig:FMD_denoising}, the single-image denoised prediction has a horizontal feature (red box) that is not present in the ground truth, and in Fig.~\ref{fig:MPM_denoising}, there is a dark region (red box) that is not present in the ground truth. These areas of hallucination have high uncertainty after single-image denoising. However, after multi-image denoising, the hallucinations disappear and the predicted features closely align with the ground truth. Concurrently, the model's uncertainty in these regions goes down substantially as they converge to the ground truth features. From this, we can see that denoisers can produce realistic-looking features that are not truly present in the sample, but our uncertainty technique can help pinpoint these regions. Furthermore, multi-image denoising can help reduce these hallucinations and increase the model's confidence in its own prediction.



\subsection{Uncertainty Bounds are Related to Noise of the Sample}
\begin{figure}[h!]
    \centering
    \includegraphics[width=\linewidth]{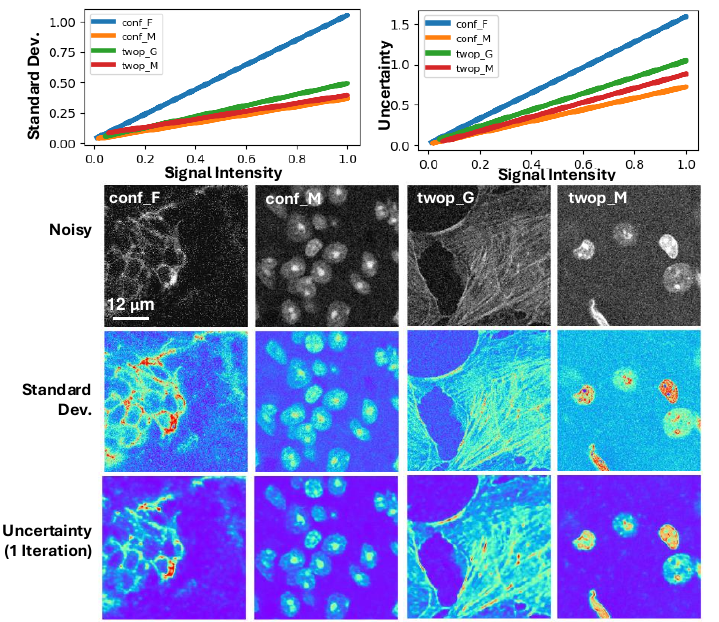}
    \captionsetup{width=1\linewidth}
    \caption{\textbf{Uncertainty vs. Noise Variance}: (top) We plot a best-fit line of the noise standard deviation (left) and the uncertainty (right) as a function of the signal intensity for several images from the FMD dataset. Images with larger slopes have a larger noise variance. This coincides with our estimated single-image uncertainty prediction, which predicts a higher uncertainty for noisier images. (bottom) The corresponding FMD images, their standard deviation from 20 noisy measurements, and the predicted single-image uncertainty.}
    \label{fig:unc_range}
\end{figure}

Next, we analyze our predicted uncertainty and compare this to the variance of the noisy measurements.  Intuitively, we expect that our network should predict a larger uncertainty interval for noisier measurements since noisier measurements have a larger variance. In addition, microscopy images often have signal-dependent Poisson noise, which has a higher variance for larger signal intensities. Thus, we expect our uncertainty intervals to be a function of the signal intensity. We test this in Fig.~\ref{fig:unc_range} by plotting a best-fit line of the uncertainty interval width and noise standard deviation as a function of the signal intensity for several representative images from the FMD dataset. The noise standard deviation is estimated from 20 noisy measurements, while the uncertainty is predicted from a single noisy measurement. Here we see that the standard deviation and predicted uncertainty are signal-dependent, with higher intensities having a higher standard deviation and uncertainty, respectively, Fig.~\ref{fig:unc_range}(top). Also, we can see that images with a higher standard deviation mostly have a higher predicted uncertainty. See Suppl. for more examples and failure cases where this trend does not hold.

\subsection{Adaptive Sample Acquisition and Uncertainty Informed Denoising}
\begin{figure}[h!]
    \centering
    \includegraphics[width=\linewidth]{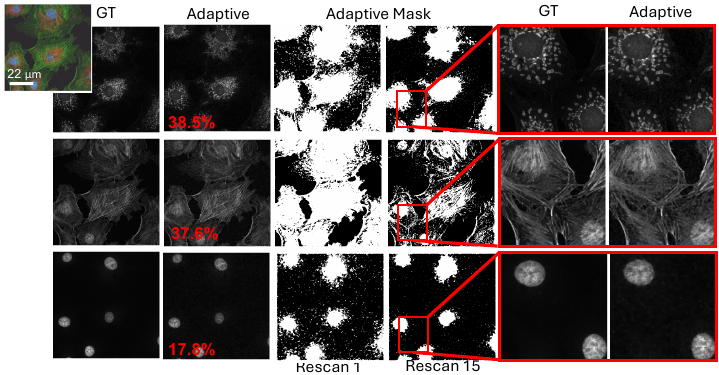}
    \captionsetup{width=1\linewidth}
    \caption{\textbf{Adaptive Acquisition of two-photon images}: We compare three different color channels from a two-photon sample from the FMD dataset. After single-image denoising, we threshold the uncertainty to obtain an adaptive mask (center).  Each channel has a different percentage of rescanned pixels (red text), which leads to a total light dosage savings of x2.6, x2.7, x5.6, respectively. With subsequent adaptive scans, the uncertainty decreases, leading to adaptive masks with fewer regions to rescan. More pixels are rescanned for dense samples, whereas fewer pixels are rescanned for sparse samples. The adaptively rescanned images retain fine features, while saving total time and light dose.}
    \label{fig:adaptive_mask_changing}
\end{figure}

Does uncertainty-guided adaptive acquisition retain image quality, while decreasing the total number of pixels needed and increasing the light-dosage savings? We investigated the effects of different adaptive percentages on different two-photon samples (Fig.~\ref{fig:adaptive_mask_changing}). In this experiment, we show the adaptive mask slowly changing through iterations- with earlier iterations needing more pixels and later iterations exclusively segmenting areas of high signal to adaptively acquire. After the 20th iteration, zoomed-in samples of the adaptively acquired and denoised image demonstrate that it maintains the same level of details for sample features, while providing a significant improvement in acquisition parameters. In Fig.~\ref{fig:adaptive_mask_changing}, we show an improvement between 2.6 and 5.6 times in light dosage while retaining image quality. 

\begin{figure}[h!]
    \centering
    \includegraphics[width=\linewidth]{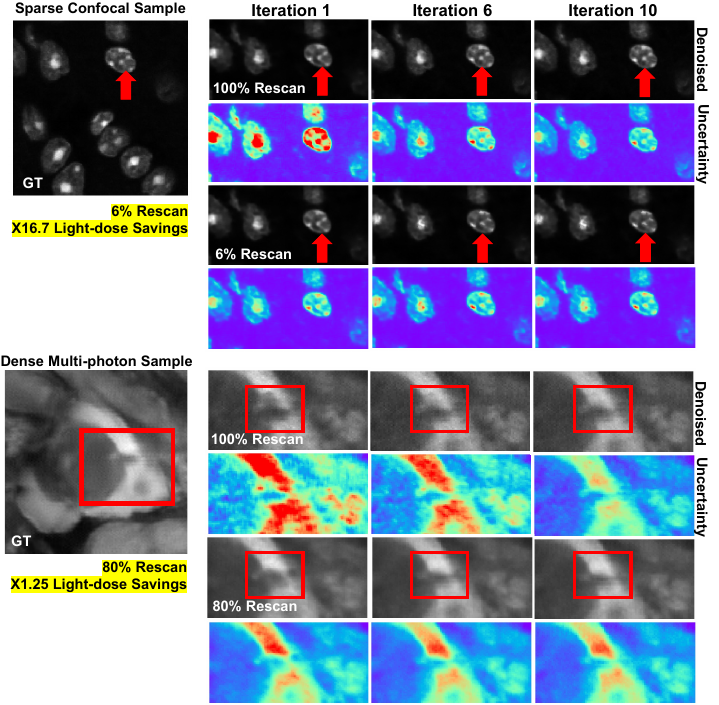}
    \captionsetup{width=1\linewidth}
    \caption{\textbf{Uncertainty Guided Adaptive Acquisition is capable of catching hallucinations}: Here we compare non-adaptive and adaptive acquisition for a confocal (top) and multiphoton (bottom) sample. The confocal sample is sparse, so thresholding the uncertainty results in rescanning only 6\% of the pixels. The multiphoton sample has higher uncertainty and requires a larger fraction of pixels to be rescanned.  In both samples, single-image denoising results in a hallucination (red arrow and red box). After both adaptive and non-adaptive multi-image denoising, the hallucinations disappear and the denoised prediction converges to the ground truth. Adaptive acquisition successfully reduces hallucinations, but has a 1.25-17$\times$ potential light dose and time savings compared to non-adaptive acquisition. }
    \label{fig:experimental_adaptive}
\end{figure}

We also show that our adaptive method is still capable of locating and resolving model hallucination in scanning microscopy images. In Fig.~\ref{fig:experimental_adaptive} we demonstrate that adaptively rescanning can resolve hallucinations in both confocal and experimental MPM samples at different rescanning percentages. In the MPM sample, rescanning at 80\% allows for a 1.25 times savings in light-dose, while in the confocal sample, rescanning at 6\% allows for a 16.7 savings in light-dose. It is important to note that optimal rescanning percentages are highly dependent on the sample itself. As shown in Fig.~\ref{fig:experimental_adaptive}, a sparse sample will need a lower percentage of pixels rescanned, while a dense sample will need a much higher rescanning percentage. 


\section{Discussion}
We presented a method to utilize learned, distribution-free uncertainty quantification for multi-image denoising and proposed an adaptive acquisition technique based on the learned uncertainty. In this paper, we demonstrate that our method of uncertainty-driven adaptive acquisition works on experimental confocal, two-photon, and multiphoton microscopy systems, showing a potential 1-16$\times$ decrease in total scanning time and light dose while successfully recovering fine structures. Our method can be adapted for different forms of scanning microscopy without explicit retraining or fine-tuning. Our network trained on FMD data can be used on different data after performing the conformal calibration step using a small calibration dataset. This is one of the advantages of conformal calibration - uncertainty predictions will still hold after calibrating for a different dataset without explicit retraining. All of the statistical guarantees will still hold; however, the size of the uncertainty bounds may increase since the network is not optimized for the data or imaging modality. We discuss this further in Suppl. Section 7. 

These speed and total light dose improvements are significant and demonstrate an important step towards faster and gentler scanning microscopy, which will enable the imaging of a new class of interesting samples and lead to new scientific insights and advances.  

Furthermore, we demonstrate how deep learning methods for microscopy can be designed to be trustworthy by building in uncertainty quantification to provide error bars for each prediction. Our method successfully identified model hallucinations, which were reduced by taking more measurements or adaptively rescanning the most uncertain regions of the sample. Our method of quantifying uncertainty provides guarantees for the reliability of the prediction. Uncertainty quantification should become standard practice when using deep-learning techniques for scientific and medical imaging to reduce hallucinations and build confidence in image predictions. We believe that the distribution-free learned uncertainty quantification presented here is an attractive path toward this due to its ease of use, fast computational time, and statistical guarantees. 

Our uncertainty-driven adaptive acquisition technique stands out from existing methods in that it is both sample and algorithm-informed. This tight coupling means that we choose to acquire new measurements not only based on the sample, but also based on how well the algorithm is performing on our measurements. Additionally, our model is generalizable in that it was trained on the FMD dataset, but demonstrates expected performance in the experimentally acquired MPM dataset. To the best of our knowledge, we are the first to propose adaptive acquisition based on algorithm uncertainty. We believe such techniques will become widespread for efficient, adaptive computational imaging systems that design their sampling scheme in an optimal, sample and algorithm-informed way.


Limitations include that we synthetically performed adaptive acquisition by capturing multiple measurements of the same sample and digitally rescanning without actual hardware rescanning. Future directions include utilizing adaptive uncertainty-driven acquisition in-the-loop with hardware using random-access scanning, such as an adaptive excitation source~\cite{li2020adaptive}, or by incorporating hardware constraints to the adaptive acquisition scheme (e.g., rescanning rows instead of random access points). Similarly, we currently only consider non-moving, 2D samples. Adapting this method for 3D, moving samples could make this more broadly applicable for neuroscience and the imaging of living animals. 

\begin{backmatter}

\bmsection{Acknowledgment} We would like to express our sincere gratitude to Ellen Kan, Keith Isaacson, Peter Movilla, and other staff from Newton Wellesley Hospital
for facilitating the tissue collection program with MIT as well as the patients who donated their tissues for our research. We thank Fan Wang, Manuel Levy, and Eva Lendaro for
providing the mouse whisker pad tissue for nonlinear imaging.  The authors acknowledge the MIT SuperCloud and Lincoln Laboratory Supercomputing Center for providing (HPC, database, consultation) resources that have contributed to the research results reported within this paper.
\bmsection{Data availability} Code will be available on a publicly accessible GitHub repository. Data underlying the FMD results presented in this paper are available in Dataset 1 \cite{mannam2019fluorescence}. Data underlying the experimental MPM results will be available on the github repository.\
\bmsection{Funding} 
We acknowledge MIT startup funds, CZI Dynamic Imaging via Chan Zuckerberg Donor Advised Fund (DAF) through the Silicon Valley Community Foundation (SVCF) 034539-00001, Massachusetts General Hospital 034902-00001, and NSF CAREER Award 035177-00001. Kristina Monakhova acknowledges funding from the MIT Postdoctoral Fellowship for Engineering Excellence. Cassandra Ye would like to thank the MIT EECS UROP and SUPERUROP programs for their continued support in funding. K.L. acknowledges support from the MIT Irwin Mark Jacobs (1957) and Joan Klein Jacobs Presidential Fellowship. 

\bmsection{Disclosures} The authors declare no conflicts of interest.

\bmsection{Supplemental Document} See Supplement 1 for supporting content.


\end{backmatter}


\bibliography{bib}

\begin{thebibliography}{10}
\newcommand{\enquote}[1]{``#1''}

\bibitem{wilson1990confocal}
T.~Wilson \emph{et~al.}, \emph{Confocal microscopy}, vol. 426 (Academic press, 1990).

\bibitem{attardo2015impermanence}
A.~Attardo, J.~E. Fitzgerald, and M.~J. Schnitzer, \enquote{Impermanence of dendritic spines in live adult ca1 hippocampus,} {\protect\JournalTitle{Nature}} \textbf{523}, 592--596 (2015).

\bibitem{lu2017video}
R.~Lu, W.~Sun, Y.~Liang, A.~Kerlin, J.~Bierfeld, J.~D. Seelig, D.~E. Wilson, B.~Scholl, B.~Mohar, M.~Tanimoto \emph{et~al.}, \enquote{Video-rate volumetric functional imaging of the brain at synaptic resolution,} {\protect\JournalTitle{Nature neuroscience}} \textbf{20}, 620--628 (2017).

\bibitem{prevedel2016fast}
R.~Prevedel, A.~J. Verhoef, A.~J. Pern{\'\i}a-Andrade, S.~Weisenburger, B.~S. Huang, T.~N{\"o}bauer, A.~Fern{\'a}ndez, J.~E. Delcour, P.~Golshani, A.~Baltuska \emph{et~al.}, \enquote{Fast volumetric calcium imaging across multiple cortical layers using sculpted light,} {\protect\JournalTitle{Nature methods}} \textbf{13}, 1021--1028 (2016).

\bibitem{perrin2020frontiers}
L.~Perrin, B.~Bayarmagnai, and B.~Gligorijevic, \enquote{Frontiers in intravital multiphoton microscopy of cancer,} {\protect\JournalTitle{Cancer Reports}} \textbf{3}, e1192 (2020).

\bibitem{you2021label}
S.~You, E.~J. Chaney, H.~Tu, Y.~Sun, S.~Sinha, and S.~A. Boppart, \enquote{Label-free deep profiling of the tumor microenvironment,} {\protect\JournalTitle{Cancer research}} \textbf{81}, 2534--2544 (2021).

\bibitem{fuentes2019second}
C.~G. Fuentes-Corona, J.~Licea-Rodriguez, R.~Younger, R.~Rangel-Rojo, E.~O. Potma, and I.~Rocha-Mendoza, \enquote{Second harmonic generation signal from type i collagen fibers grown in vitro,} {\protect\JournalTitle{Biomedical optics express}} \textbf{10}, 6449--6461 (2019).

\bibitem{debarre2006imaging}
D.~D{\'e}barre, W.~Supatto, A.-M. Pena, A.~Fabre, T.~Tordjmann, L.~Combettes, M.-C. Schanne-Klein, and E.~Beaurepaire, \enquote{Imaging lipid bodies in cells and tissues using third-harmonic generation microscopy,} {\protect\JournalTitle{Nature methods}} \textbf{3}, 47--53 (2006).

\bibitem{barad1997nonlinear}
Y.~Barad, H.~Eisenberg, M.~Horowitz, and Y.~Silberberg, \enquote{Nonlinear scanning laser microscopy by third harmonic generation,} {\protect\JournalTitle{Applied Physics Letters}} \textbf{70}, 922--924 (1997).

\bibitem{zipfel2003live}
W.~R. Zipfel, R.~M. Williams, R.~Christie, A.~Y. Nikitin, B.~T. Hyman, and W.~W. Webb, \enquote{Live tissue intrinsic emission microscopy using multiphoton-excited native fluorescence and second harmonic generation,} {\protect\JournalTitle{Proceedings of the National Academy of Sciences}} \textbf{100}, 7075--7080 (2003).

\bibitem{liu2024deep}
K.~Liu, H.~Cao, K.~Shashaty, L.-Y. Yu, S.~Spitz, F.~M. Pramotton, Z.~Wan, E.~L. Kan, E.~N. Tevonian, M.~Levy \emph{et~al.}, \enquote{Deep and dynamic metabolic and structural imaging in living tissues,} {\protect\JournalTitle{Science Advances}} \textbf{10}, eadp2438 (2024).

\bibitem{periasamy2020special}
A.~Periasamy, K.~K{\"o}nig, and P.~So, \enquote{Special section guest editorial: Thirty years of multiphoton microscopy in the biomedical sciences,} {\protect\JournalTitle{Journal of Biomedical Optics}} \textbf{25} (2020).

\bibitem{CHEN20111001}
A.~C.-H. Chen, C.~McNeilly, A.~P.-Y. Liu, C.~J. Flaim, L.~Cuttle, M.~Kendall, R.~M. Kimble, H.~Shimizu, and J.~R. McMillan, \enquote{Second harmonic generation and multiphoton microscopic detection of collagen without the need for species specific antibodies,} {\protect\JournalTitle{Burns}} \textbf{37}, 1001--1009 (2011).

\bibitem{skala2007vivo}
M.~C. Skala, K.~M. Riching, A.~Gendron-Fitzpatrick, J.~Eickhoff, K.~W. Eliceiri, J.~G. White, and N.~Ramanujam, \enquote{In vivo multiphoton microscopy of nadh and fad redox states, fluorescence lifetimes, and cellular morphology in precancerous epithelia,} {\protect\JournalTitle{Proceedings of the National Academy of Sciences}} \textbf{104}, 19494--19499 (2007).

\bibitem{brown2001vivo}
E.~B. Brown, R.~B. Campbell, Y.~Tsuzuki, L.~Xu, P.~Carmeliet, D.~Fukumura, and R.~K. Jain, \enquote{In vivo measurement of gene expression, angiogenesis and physiological function in tumors using multiphoton laser scanning microscopy,} {\protect\JournalTitle{Nature medicine}} \textbf{7}, 864--868 (2001).

\bibitem{you2018intravital}
S.~You, H.~Tu, E.~J. Chaney, Y.~Sun, Y.~Zhao, A.~J. Bower, Y.-Z. Liu, M.~Marjanovic, S.~Sinha, Y.~Pu \emph{et~al.}, \enquote{Intravital imaging by simultaneous label-free autofluorescence-multiharmonic microscopy,} {\protect\JournalTitle{Nature communications}} \textbf{9}, 2125 (2018).

\bibitem{volpe2023roadmap}
G.~Volpe, C.~W{\"a}hlby, L.~Tian, M.~Hecht, A.~Yakimovich, K.~Monakhova, L.~Waller, I.~F. Sbalzarini, C.~A. Metzler, M.~Xie \emph{et~al.}, \enquote{Roadmap on deep learning for microscopy,} {\protect\JournalTitle{ArXiv}}  (2023).

\bibitem{weigert2018content}
M.~Weigert, U.~Schmidt, T.~Boothe, A.~M{\"u}ller, A.~Dibrov, A.~Jain, B.~Wilhelm, D.~Schmidt, C.~Broaddus, S.~Culley \emph{et~al.}, \enquote{Content-aware image restoration: pushing the limits of fluorescence microscopy,} {\protect\JournalTitle{Nature methods}} \textbf{15}, 1090--1097 (2018).

\bibitem{platisa2023high}
J.~Platisa, X.~Ye, A.~M. Ahrens, C.~Liu, I.~A. Chen, I.~G. Davison, L.~Tian, V.~A. Pieribone, and J.~L. Chen, \enquote{High-speed low-light in vivo two-photon voltage imaging of large neuronal populations,} {\protect\JournalTitle{Nature Methods}} pp. 1--9 (2023).

\bibitem{von2021democratising}
L.~von Chamier, R.~F. Laine, J.~Jukkala, C.~Spahn, D.~Krentzel, E.~Nehme, M.~Lerche, S.~Hern{\'a}ndez-P{\'e}rez, P.~K. Mattila, E.~Karinou \emph{et~al.}, \enquote{Democratising deep learning for microscopy with zerocostdl4mic,} {\protect\JournalTitle{Nature communications}} \textbf{12}, 2276 (2021).

\bibitem{hsu2022three}
C.-W. Hsu, C.-Y. Lin, Y.~Y. Hu, C.-Y. Wang, S.-T. Chang, A.-S. Chiang, and S.-J. Chen, \enquote{Three-dimensional-generator u-net for dual-resonant scanning multiphoton microscopy image inpainting and denoising,} {\protect\JournalTitle{Biomedical Optics Express}} \textbf{13}, 6273--6283 (2022).

\bibitem{zhang2019poisson}
Y.~Zhang, Y.~Zhu, E.~Nichols, Q.~Wang, S.~Zhang, C.~Smith, and S.~Howard, \enquote{A poisson-gaussian denoising dataset with real fluorescence microscopy images,} in \emph{Proceedings of the IEEE/CVF Conference on Computer Vision and Pattern Recognition,}  (2019), pp. 11710--11718.

\bibitem{lee2020mu}
S.~Lee, M.~Negishi, H.~Urakubo, H.~Kasai, and S.~Ishii, \enquote{Mu-net: Multi-scale u-net for two-photon microscopy image denoising and restoration,} {\protect\JournalTitle{Neural Networks}} \textbf{125}, 92--103 (2020).

\bibitem{laine2021imaging}
R.~F. Laine, G.~Jacquemet, and A.~Krull, \enquote{Imaging in focus: an introduction to denoising bioimages in the era of deep learning,} {\protect\JournalTitle{The international journal of biochemistry \& cell biology}} \textbf{140}, 106077 (2021).

\bibitem{manifold2019denoising}
B.~Manifold, E.~Thomas, A.~T. Francis, A.~H. Hill, and D.~Fu, \enquote{Denoising of stimulated raman scattering microscopy images via deep learning,} {\protect\JournalTitle{Biomedical optics express}} \textbf{10}, 3860--3874 (2019).

\bibitem{antun2020instabilities}
V.~Antun, F.~Renna, C.~Poon, B.~Adcock, and A.~C. Hansen, \enquote{On instabilities of deep learning in image reconstruction and the potential costs of ai,} {\protect\JournalTitle{Proceedings of the National Academy of Sciences}} \textbf{117}, 30088--30095 (2020).

\bibitem{ji2023survey}
Z.~Ji, N.~Lee, R.~Frieske, T.~Yu, D.~Su, Y.~Xu, E.~Ishii, Y.~J. Bang, A.~Madotto, and P.~Fung, \enquote{Survey of hallucination in natural language generation,} {\protect\JournalTitle{ACM Computing Surveys}} \textbf{55}, 1--38 (2023).

\bibitem{helou2020stochastic}
M.~E. Helou, R.~Zhou, and S.~Süsstrunk, \enquote{Stochastic frequency masking to improve super-resolution and denoising networks,}  (2020).

\bibitem{bhadra2021hallucinations}
S.~Bhadra, V.~A. Kelkar, F.~J. Brooks, and M.~A. Anastasio, \enquote{On hallucinations in tomographic image reconstruction,} {\protect\JournalTitle{IEEE transactions on medical imaging}} \textbf{40}, 3249--3260 (2021).

\bibitem{gal2016dropout}
Y.~Gal and Z.~Ghahramani, \enquote{Dropout as a bayesian approximation: Representing model uncertainty in deep learning,} in \emph{international conference on machine learning,}  (PMLR, 2016), pp. 1050--1059.

\bibitem{abdar2021review}
M.~Abdar, F.~Pourpanah, S.~Hussain, D.~Rezazadegan, L.~Liu, M.~Ghavamzadeh, P.~Fieguth, X.~Cao, A.~Khosravi, U.~R. Acharya \emph{et~al.}, \enquote{A review of uncertainty quantification in deep learning: Techniques, applications and challenges,} {\protect\JournalTitle{Information fusion}} \textbf{76}, 243--297 (2021).

\bibitem{Xue_2019}
Y.~Xue, S.~Cheng, Y.~Li, and L.~Tian, \enquote{Reliable deep-learning-based phase imaging with uncertainty quantification,} {\protect\JournalTitle{Optica}} \textbf{6}, 618 (2019).

\bibitem{angelopoulos2022image}
A.~N. Angelopoulos, A.~P. Kohli, S.~Bates, M.~Jordan, J.~Malik, T.~Alshaabi, S.~Upadhyayula, and Y.~Romano, \enquote{Image-to-image regression with distribution-free uncertainty quantification and applications in imaging,} in \emph{International Conference on Machine Learning,}  (PMLR, 2022), pp. 717--730.

\bibitem{koenker1978regression}
R.~Koenker and G.~Bassett~Jr, \enquote{Regression quantiles,} {\protect\JournalTitle{Econometrica: journal of the Econometric Society}} pp. 33--50 (1978).

\bibitem{angelopoulos2022conformal}
A.~N. Angelopoulos, S.~Bates, A.~Fisch, L.~Lei, and T.~Schuster, \enquote{Conformal risk control,} {\protect\JournalTitle{arXiv preprint arXiv:2208.02814}}  (2022).

\bibitem{portilla2003image}
J.~Portilla, V.~Strela, M.~J. Wainwright, and E.~P. Simoncelli, \enquote{Image denoising using scale mixtures of gaussians in the wavelet domain,} {\protect\JournalTitle{IEEE Transactions on Image processing}} \textbf{12}, 1338--1351 (2003).

\bibitem{dabov2007image}
K.~Dabov, A.~Foi, V.~Katkovnik, and K.~Egiazarian, \enquote{Image denoising by sparse 3-d transform-domain collaborative filtering,} {\protect\JournalTitle{IEEE Transactions on image processing}} \textbf{16}, 2080--2095 (2007).

\bibitem{zhang2017beyond}
K.~Zhang, W.~Zuo, Y.~Chen, D.~Meng, and L.~Zhang, \enquote{Beyond a gaussian denoiser: Residual learning of deep cnn for image denoising,} {\protect\JournalTitle{IEEE transactions on image processing}} \textbf{26}, 3142--3155 (2017).

\bibitem{gottschling2020troublesome}
N.~M. Gottschling, V.~Antun, B.~Adcock, and A.~C. Hansen, \enquote{The troublesome kernel: why deep learning for inverse problems is typically unstable,} {\protect\JournalTitle{arXiv preprint arXiv:2001.01258}}  (2020).

\bibitem{blau2018perception}
Y.~Blau and T.~Michaeli, \enquote{The perception-distortion tradeoff,} in \emph{Proceedings of the IEEE conference on computer vision and pattern recognition,}  (2018), pp. 6228--6237.

\bibitem{prakash2020fully2}
M.~Prakash, A.~Krull, and F.~Jug, \enquote{Fully unsupervised diversity denoising with convolutional variational autoencoders,} {\protect\JournalTitle{arXiv preprint arXiv:2006.06072}}  (2020).

\bibitem{prakash2021interpretable}
M.~Prakash, M.~Delbracio, P.~Milanfar, and F.~Jug, \enquote{Interpretable unsupervised diversity denoising and artefact removal,} {\protect\JournalTitle{arXiv preprint arXiv:2104.01374}}  (2021).

\bibitem{liba2019handheld}
O.~Liba, K.~Murthy, Y.-T. Tsai, T.~Brooks, T.~Xue, N.~Karnad, Q.~He, J.~T. Barron, D.~Sharlet, R.~Geiss \emph{et~al.}, \enquote{Handheld mobile photography in very low light,} {\protect\JournalTitle{ACM Transactions on Graphics (TOG)}} \textbf{38}, 1--16 (2019).

\bibitem{monakhova2022dancing}
K.~Monakhova, S.~R. Richter, L.~Waller, and V.~Koltun, \enquote{Dancing under the stars: video denoising in starlight,} in \emph{Proceedings of the IEEE/CVF Conference on Computer Vision and Pattern Recognition,}  (2022), pp. 16241--16251.

\bibitem{batson2019noise2self}
J.~Batson and L.~Royer, \enquote{Noise2self: Blind denoising by self-supervision,} in \emph{International Conference on Machine Learning,}  (PMLR, 2019), pp. 524--533.

\bibitem{krull2019noise2void}
A.~Krull, T.-O. Buchholz, and F.~Jug, \enquote{Noise2void-learning denoising from single noisy images,} in \emph{Proceedings of the IEEE/CVF conference on computer vision and pattern recognition,}  (2019), pp. 2129--2137.

\bibitem{lecoq2021removing}
J.~Lecoq, M.~Oliver, J.~H. Siegle, N.~Orlova, P.~Ledochowitsch, and C.~Koch, \enquote{Removing independent noise in systems neuroscience data using deepinterpolation,} {\protect\JournalTitle{Nature methods}} \textbf{18}, 1401--1408 (2021).

\bibitem{li2021reinforcing}
X.~Li, G.~Zhang, J.~Wu, Y.~Zhang, Z.~Zhao, X.~Lin, H.~Qiao, H.~Xie, H.~Wang, L.~Fang \emph{et~al.}, \enquote{Reinforcing neuron extraction and spike inference in calcium imaging using deep self-supervised denoising,} {\protect\JournalTitle{Nature methods}} \textbf{18}, 1395--1400 (2021).

\bibitem{gal2022bayesian}
Y.~Gal, P.~Koumoutsakos, F.~Lanusse, G.~Louppe, and C.~Papadimitriou, \enquote{Bayesian uncertainty quantification for machine-learned models in physics,} {\protect\JournalTitle{Nature Reviews Physics}} \textbf{4}, 573--577 (2022).

\bibitem{zhang2021blindnet}
X.~Zhang, F.~Wang, and G.~Situ, \enquote{Blindnet: an untrained learning approach toward computational imaging with model uncertainty,} {\protect\JournalTitle{Journal of Physics D: Applied Physics}} \textbf{55}, 034001 (2021).

\bibitem{kendall2017uncertainties}
A.~Kendall and Y.~Gal, \enquote{What uncertainties do we need in bayesian deep learning for computer vision?} {\protect\JournalTitle{Advances in neural information processing systems}} \textbf{30} (2017).

\bibitem{neal2012bayesian}
R.~M. Neal, \emph{Bayesian learning for neural networks}, vol. 118 (Springer Science \& Business Media, 2012).

\bibitem{shang2023approximating}
R.~Shang, M.~A. O’Brien, F.~Wang, G.~Situ, and G.~P. Luke, \enquote{Approximating the uncertainty of deep learning reconstruction predictions in single-pixel imaging,} {\protect\JournalTitle{Communications engineering}} \textbf{2}, 53 (2023).

\bibitem{NIPS2017_9ef2ed4b}
B.~Lakshminarayanan, A.~Pritzel, and C.~Blundell, \enquote{Simple and scalable predictive uncertainty estimation using deep ensembles,} in \emph{Advances in Neural Information Processing Systems,}  vol.~30 I.~Guyon, U.~V. Luxburg, S.~Bengio, H.~Wallach, R.~Fergus, S.~Vishwanathan, and R.~Garnett, eds. (Curran Associates, Inc., 2017).

\bibitem{hirsh2022sparsifying}
S.~M. Hirsh, D.~A. Barajas-Solano, and J.~N. Kutz, \enquote{Sparsifying priors for bayesian uncertainty quantification in model discovery,} {\protect\JournalTitle{Royal Society Open Science}} \textbf{9}, 211823 (2022).

\bibitem{freeman2015improving}
L.~J. Freeman and K.~Fronczyk, \enquote{Improving reliability estimates with bayesian statistics,} {\protect\JournalTitle{IDA Document NS D-5452}}  (2015).

\bibitem{xue2019reliable}
Y.~Xue, S.~Cheng, Y.~Li, and L.~Tian, \enquote{Reliable deep-learning-based phase imaging with uncertainty quantification,} {\protect\JournalTitle{Optica}} \textbf{6}, 618--629 (2019).

\bibitem{angelopoulos2021gentle}
A.~N. Angelopoulos and S.~Bates, \enquote{A gentle introduction to conformal prediction and distribution-free uncertainty quantification,} {\protect\JournalTitle{arXiv preprint arXiv:2107.07511}}  (2021).

\bibitem{teneggi2023trust}
J.~Teneggi, M.~Tivnan, W.~Stayman, and J.~Sulam, \enquote{How to trust your diffusion model: A convex optimization approach to conformal risk control,} in \emph{International Conference on Machine Learning,}  (PMLR, 2023), pp. 33940--33960.

\bibitem{cheung2024metric}
M.~Y. Cheung, T.~J. Netherton, L.~E. Court, A.~Veeraraghavan, and G.~Balakrishnan, \enquote{Metric-guided image reconstruction bounds via conformal prediction,} {\protect\JournalTitle{ArXiv}}  (2024).

\bibitem{connolly1985determination}
C.~Connolly, \enquote{The determination of next best views,} in \emph{Proceedings. 1985 IEEE international conference on robotics and automation,}  vol.~2 (IEEE, 1985), pp. 432--435.

\bibitem{aloimonos1988active}
J.~Aloimonos, I.~Weiss, and A.~Bandyopadhyay, \enquote{Active vision,} {\protect\JournalTitle{International journal of computer vision}} \textbf{1}, 333--356 (1988).

\bibitem{maver1993occlusions}
J.~Maver and R.~Bajcsy, \enquote{Occlusions as a guide for planning the next view,} {\protect\JournalTitle{IEEE transactions on pattern analysis and machine intelligence}} \textbf{15}, 417--433 (1993).

\bibitem{dunn2009developing}
E.~Dunn, J.~Van Den~Berg, and J.-M. Frahm, \enquote{Developing visual sensing strategies through next best view planning,} in \emph{2009 IEEE/RSJ International Conference on Intelligent Robots and Systems,}  (IEEE, 2009), pp. 4001--4008.

\bibitem{sheinin2016next}
M.~Sheinin and Y.~Y. Schechner, \enquote{The next best underwater view,} in \emph{Proceedings of the IEEE conference on computer vision and pattern recognition,}  (2016), pp. 3764--3773.

\bibitem{lekeufack2024conformal}
J.~Lekeufack, A.~N. Angelopoulos, A.~Bajcsy, M.~I. Jordan, and J.~Malik, \enquote{Conformal decision theory: Safe autonomous decisions from imperfect predictions,} in \emph{2024 IEEE International Conference on Robotics and Automation (ICRA),}  (IEEE, 2024), pp. 11668--11675.

\bibitem{hoebe2007controlled}
R.~Hoebe, C.~Van~Oven, T.~Gadella~Jr, P.~Dhonukshe, C.~Van~Noorden, and E.~Manders, \enquote{Controlled light-exposure microscopy reduces photobleaching and phototoxicity in fluorescence live-cell imaging,} {\protect\JournalTitle{Nature biotechnology}} \textbf{25}, 249--253 (2007).

\bibitem{abouakil2021adaptive}
F.~Abouakil, H.~Meng, M.-A. Burcklen, H.~Rigneault, F.~Galland, and L.~LeGoff, \enquote{An adaptive microscope for the imaging of biological surfaces,} {\protect\JournalTitle{Light: Science \& Applications}} \textbf{10}, 210 (2021).

\bibitem{dreier2019smart}
J.~Dreier, M.~Castello, G.~Coceano, R.~C{\'a}ceres, J.~Plastino, G.~Vicidomini, and I.~Testa, \enquote{Smart scanning for low-illumination and fast resolft nanoscopy in vivo,} {\protect\JournalTitle{Nature communications}} \textbf{10}, 556 (2019).

\bibitem{chu2007enhanced}
K.~K. Chu, D.~Lim, and J.~Mertz, \enquote{Enhanced weak-signal sensitivity in two-photon microscopy by adaptive illumination,} {\protect\JournalTitle{Optics letters}} \textbf{32}, 2846--2848 (2007).

\bibitem{li2020adaptive}
B.~Li, C.~Wu, M.~Wang, K.~Charan, and C.~Xu, \enquote{An adaptive excitation source for high-speed multiphoton microscopy,} {\protect\JournalTitle{Nature methods}} \textbf{17}, 163--166 (2020).

\bibitem{royer2016adaptive}
L.~A. Royer, W.~C. Lemon, R.~K. Chhetri, Y.~Wan, M.~Coleman, E.~W. Myers, and P.~J. Keller, \enquote{Adaptive light-sheet microscopy for long-term, high-resolution imaging in living organisms,} {\protect\JournalTitle{Nature biotechnology}} \textbf{34}, 1267--1278 (2016).

\bibitem{mcdole2018toto}
K.~McDole, L.~Guignard, F.~Amat, A.~Berger, G.~Malandain, L.~A. Royer, S.~C. Turaga, K.~Branson, and P.~J. Keller, \enquote{In toto imaging and reconstruction of post-implantation mouse development at the single-cell level,} {\protect\JournalTitle{Cell}} \textbf{175}, 859--876 (2018).

\bibitem{alvelid2022event}
J.~Alvelid, M.~Damenti, C.~Sgattoni, and I.~Testa, \enquote{Event-triggered sted imaging,} {\protect\JournalTitle{Nature Methods}} \textbf{19}, 1268--1275 (2022).

\bibitem{mahecic2022event}
D.~Mahecic, W.~L. Stepp, C.~Zhang, J.~Griffi{\'e}, M.~Weigert, and S.~Manley, \enquote{Event-driven acquisition for content-enriched microscopy,} {\protect\JournalTitle{Nature Methods}} \textbf{19}, 1262--1267 (2022).

\bibitem{pinkard2021learned}
H.~Pinkard, H.~Baghdassarian, A.~Mujal, E.~Roberts, K.~H. Hu, D.~H. Friedman, I.~Malenica, T.~Shagam, A.~Fries, K.~Corbin \emph{et~al.}, \enquote{Learned adaptive multiphoton illumination microscopy for large-scale immune response imaging,} {\protect\JournalTitle{Nature communications}} \textbf{12}, 1916 (2021).

\bibitem{chen2022simple}
L.~Chen, X.~Chu, X.~Zhang, and J.~Sun, \enquote{Simple baselines for image restoration,} in \emph{European conference on computer vision,}  (Springer, 2022), pp. 17--33.

\bibitem{mannam2019fluorescence}
V.~Mannam, Y.~Zhang, Y.~Zhu, and S.~Howard, \enquote{Fluorescence microscopy denoising (fmd) dataset,} Tech. rep., University of Notre Dame, IN (United States) (2019).

\end{thebibliography}

\end{document}


\title{Supplement to Learned, Uncertainty-driven Adaptive Acquisition for Photon-Efficient Scanning Microscopy}

\author{Cassandra Tong Ye\authormark{1}, Jiashu Han\authormark{2}, Kunzan Liu\authormark{1}, Anastasios Angelopoulos\authormark{3}, Linda Griffith\authormark{4}, 
Kristina Monakhova\authormark{1*}, Sixian You\authormark{1*}}

\address{\authormark{1}Research Laboratory of Electronics (RLE) in the Department of Electrical Engineering and Computer Science, Massachusetts Institute of Technology. Cambridge, MA USA 02139\\
\authormark{2}Fu Foundation School of Engineering and Applied Science, Columbia University in the City of New York. New York, NY USA 10027\\
\authormark{3}Department of Electrical Engineering and Computer Science, University of California, Berkeley. Berkeley, CA USA 94720\\
\authormark{4}Department of Biological Engineering, Massachusetts Institute of Technology. Cambridge, MA USA 02139}

\email{\authormark{*}Sixian You: sixian@mit.edu (617-253-4600)\\\authormark{*}Kristina Monakhova: monakhova@cornell.edu}

\section{Network comparisons}
\begin{figure}
    \centering
    \includegraphics[width=\linewidth]{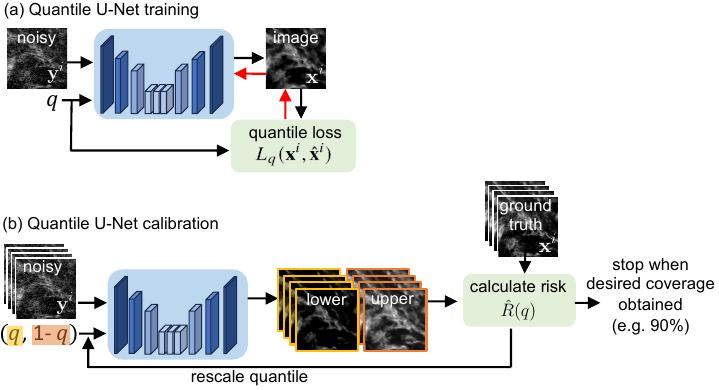}
    \captionsetup{width=1\linewidth}
    \caption{\textbf{Quantile U-Net}. (a) The quantile U-Net has a control input to adjust the quantile, which globally adjusts the predicted image and affects the loss function. During training, quantile ranges between [0,1] are randomly selected, thus training the network to predict any arbitrary quantile image rather than a specific lower and upper quantile. (b) During calibration using conformal risk control, two forward passes are needed to predict the upper bound (1-$q$) and lower bound ($q$) for each noisy image. These bounds are compared with the ground truth to determine coverage through the risk function. The quantile control input can then be updated to increase or decrease the bounds until the desired coverage is obtained. }
    \label{supp:quantile}
\end{figure}

To demonstrate the role that network architecture has on denoising and uncertainty prediction performance, we compare the performance of two networks trained on the same dataset. We compare the NAFNet architecture described in the main text to a U-Net with a control input, which we will refer to as a quantile U-Net, Fig.~\ref{supp:quantile}. Both networks are trained and calibrated using the same datasets. 

For the quantile U-Net, we use a standard attention-based U-Net used in diffusion models based on~\cite{dhariwal2021diffusion}, but replace the time embedding with quantile embedding to control the predicted quantile. The U-Net has a 256 image size and one output. As opposed to NAFNet which has three output images, the quantile U-Net has only one output image and requires three forward passes with a varying control input (e.g. $q=0.1,0.5,0.9$) to generate the lower bound, image prediction, and upper bound. During training, the control input, $q$, is randomly sampled between 0 and 1. The control input is fed into both the network and the loss function to adjust the quantile loss as a function of $q$. Thus, the network is trained to predict any arbitrary quantile rather a pre-defined quantile. Training has the following loss function:

\begin{equation}
\min_{\theta} \sum_i  L_{q_r}(\mathbf x^i,\hat{\mathbf x}^i),
\end{equation}

\noindent where $q_r$ is the quantile control input randomly sampled throughout training. 

During conformal calibration, rather than scaling the bounds with a constant, $\hat{\lambda}$, the control input, $q$ can be adjusted until the bounds guarantee the appropriate coverage, Fig.~\ref{supp:quantile}(b). For example, we might start with $q_l=0.1, q_h=1-0.1$ for the lower and upper bound images. If 90\% coverage is not achieved on the calibration dataset, these control inputs would adapted until the coverage is satisfied, for example by decreasing the value to $q_l=0.01, q_h=1-0.01$. This means that a non-uniform scaling could be applied to increase and decrease the bounds to obtain the appropriate coverage instead of the uniform scaling used for NAFNet. 

The following sections show results from both NAFNet and the quantile U-Net. Interestingly, both networks often produce similar hallucinations, which may suggest that the hallucinations are caused by data uncertainty rather than network uncertainty. Overall, the quantile U-Net has slightly sharper features for single-image denoising than NAFNet, but both converge to a similar prediction after multi-image denoising. 

\section{How prevalent are hallucinations in denoising?}
\begin{figure}
    \centering
    \includegraphics[width=\linewidth]{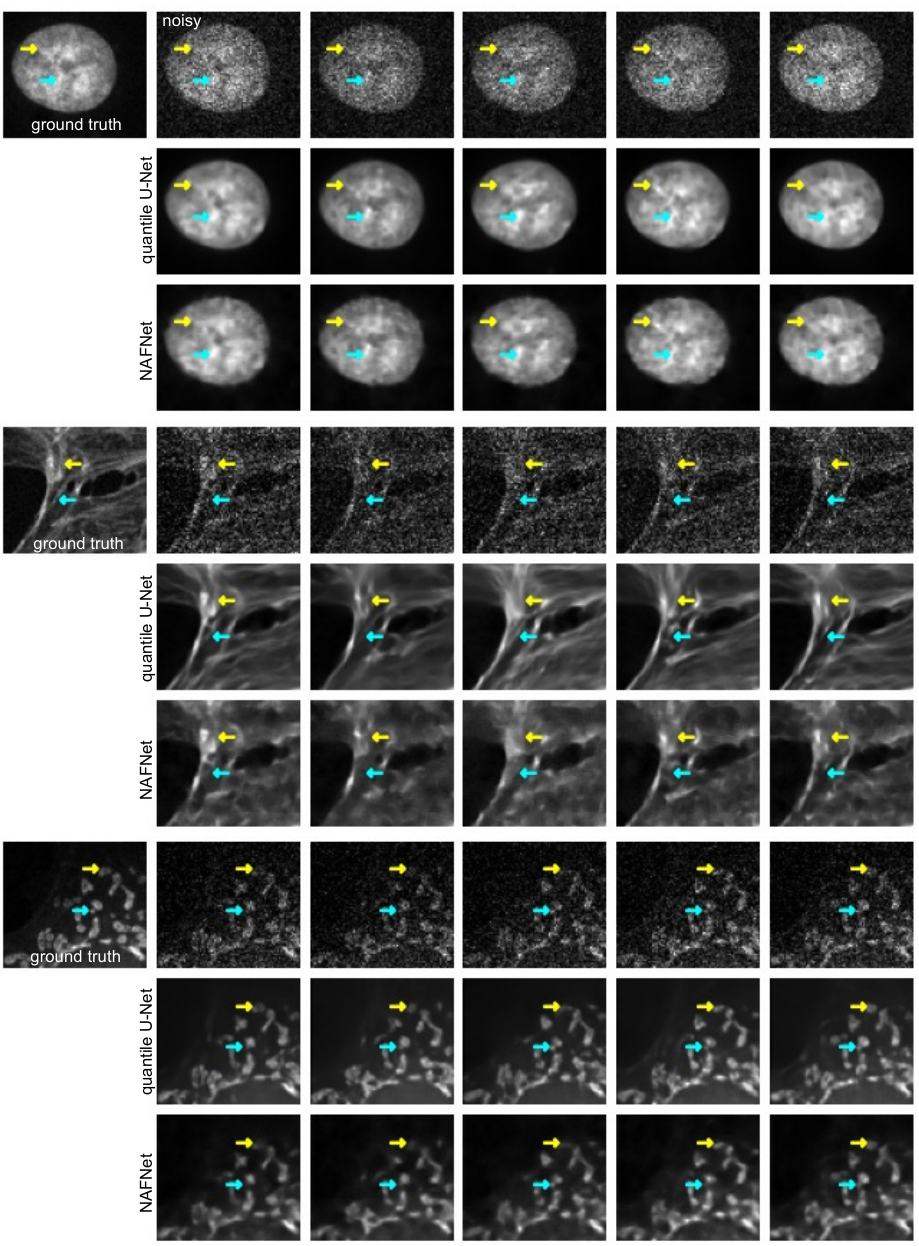}
    \captionsetup{width=1\linewidth}
    \caption{\textbf{Single-image denoising hallucinations}. Given a noisy measurement, a denoising network may infer different features and structures from the measurement. Here, we show how single-image denoising predictions vary when using a different measurement of the same sample, since the noise is different between each measurement. Moving left to right, different features and structures are present between the denoised predictions.}
    \label{supp:hallucin_single}
\end{figure}

While one might expect hallucinations to occur when solving inverse problems with a non-identity $\mathbf A$ operator, such as compressive MRI, we find that hallucinations can be quite prevalent in denoising when there is substantial noise. To test and better understand the prevalence of hallucinations for single-image denoising, we passed in measurements with the same signal, but different instances of noise to our pre-trained networks. As we can see in Fig~\ref{supp:hallucin_single}, each instance of single-image denoising produces a different prediction, sometimes with very different features present. We show the single-image denoising results for both NAFNet and the quantile U-Net for three samples. Looking at the yellow and cyan arrows in each sample, we can see that each denoised image moving left to right has different features. Each feature appears compelling, but we cannot know which is the true signal without acquiring more measurements. 

\section{Additional hallucination results}
\begin{figure}
    \centering
    \includegraphics[width=\linewidth]{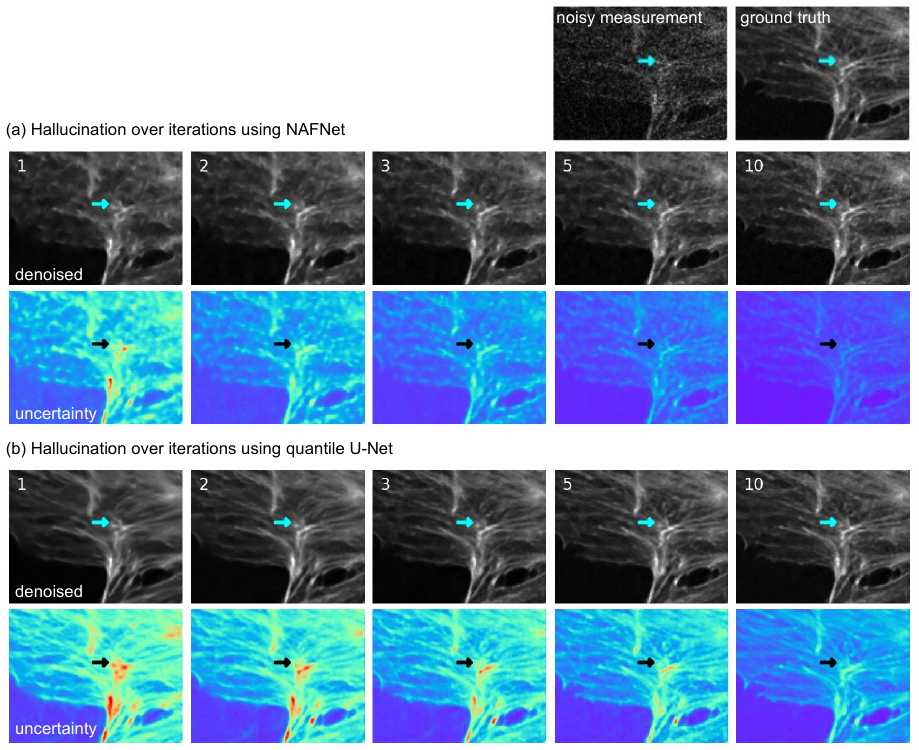}
    \captionsetup{width=1\linewidth}
    \caption{\textbf{Additional hallucinations}: Two-photon BPAE data. We compare denoising results for NAFNet \textbf{(a)} and the quantile U-Net \textbf{(b)}, showing 1,2,3,5, and 10-image denoising (left to right). . In both cases, we observe a loop structure (indicated by the arrow) that's predicted after single-image denoising, but is not present in the ground truth image. This hallucination disappears as more measurements are used for denoising. For both networks, we can see that the uncertainty is high where the hallucination is present, and the uncertainty in this region decreases as more measurements are used. } 
    \label{supp:hallucin}
\end{figure}

\begin{figure}
    \centering
    \includegraphics[width=\linewidth]{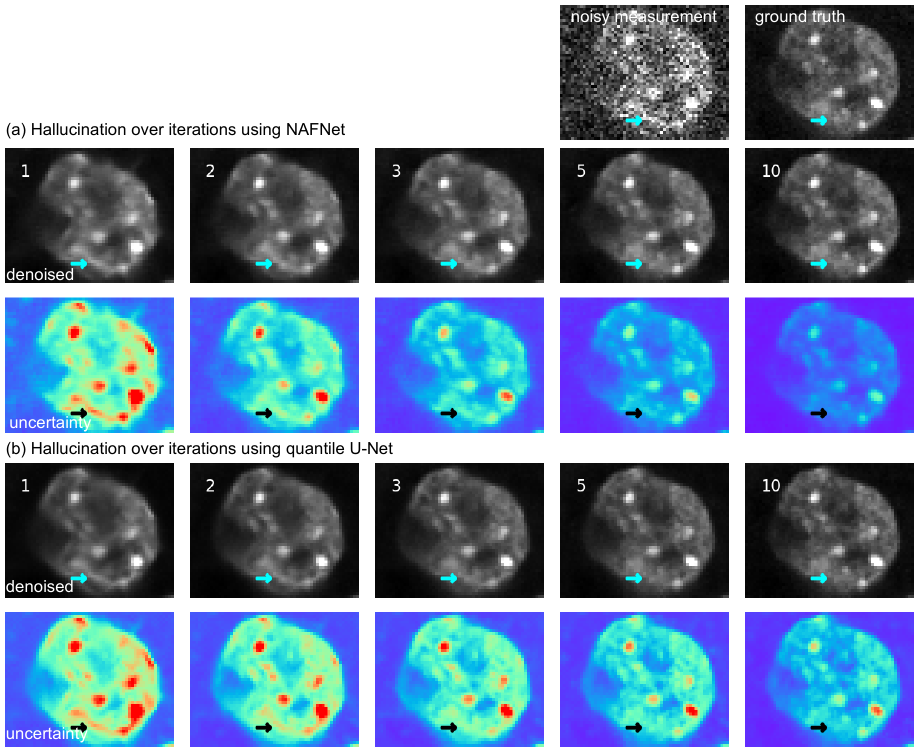}
    \captionsetup{width=1\linewidth}
    \caption{\textbf{Additional hallucinations}: Two-photon MICE data. We compare the multi-image denoising results for NAFNet \textbf{(a)}, and a quantile U-Net \textbf{(b)}, showing 1,2,3,5, and 10-image denoising (left to right).  In both cases, we see a line connecting two spots in the specimen (arrow) which is not present in the ground truth. As more measurements are used for denoising, this hallucination slowly disappears and the predictions converge to the ground truth image. This hallucination has high uncertainty at first, but the uncertainty decreases once more measurements are used.}
    \label{supp:hallucin_2pmice}
\end{figure}

\begin{figure}
    \centering
    \includegraphics[width=\linewidth]{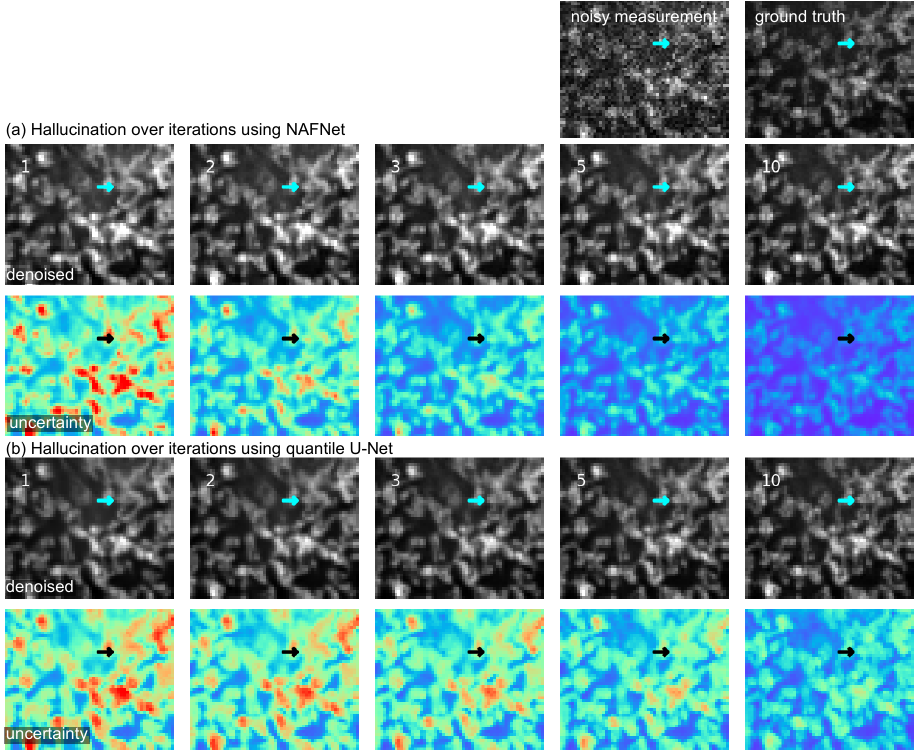}
    \captionsetup{width=1\linewidth}
    \caption{\textbf{Additional hallucinations}: Confocal BPAE data. As before, we compare the multi-image denoising results for NAFNet \textbf{(a)}, and a quantile U-Net \textbf{(b)}, showing 1,2,3,5, and 10-image denoising (left to right). This time, both networks predict a loop-feature (arrow) that is not present in the ground truth . This region has high uncertainty, but the uncertainty decreases and the hallucination disappears once more measurements are used.}
    \label{supp:hallucin_confocal}
\end{figure}

Next, we show several examples of hallucinations that are present after single-image denoising, but disappear with multi-image denoising. We compare the denoised image and uncertainty predictions obtained from NAFNet and a quantile U-Net to show how predictions and uncertainty maps vary across different networks. We show hallucinations in a two-photon BPAE sample (Fig.~\ref{supp:hallucin}), a two-photon MICE sample (Fig.~\ref{supp:hallucin_2pmice}), and a confocal BPAE sample (Fig.~\ref{supp:hallucin_confocal}). In the two-photon sample, Fig.~\ref{supp:hallucin}, we can see a prominent loop feature (cyan arrow) that is present after single-image denoising, but it not present in the ground truth. This feature has high uncertainty, and disappears after multi-image denoising with 3, 5, or 10 images. For the two-photon MICE sample, Fig.~\ref{supp:hallucin_2pmice}, there is a prominent curvy line (cyan arrow) that appears in the single-image denoised prediction for both networks that is not present in the ground truth. This region has high uncertainty, and is gone once 5 or 10 measurements are used for denoising. Finally, for the confocal BPAE sample, Fig.~\ref{supp:hallucin_confocal}, there is a prominent loop (cyan arrow) that appears in both networks after single-image denoising that is not present in the ground truth. This feature has high uncertainty and disappears after multi-image denoising. 

Interestingly, both NAFNet and the quantile U-Net produced similar hallucinations in each example, although they have slightly different predictions and uncertainty maps. The quantile U-Net tends to predict smooth curves in single-image denoising, while NAFNet produces more irregular shapes (Fig.~\ref{supp:hallucin}(left)). This suggests that the noise structure in the measurements in contributing to the perceived hallucinations, and with multiple measurements, incorrectly perceived structures vanish into the noise. Both networks converge to a similar prediction after multi-image denoising.

\section{Comparison of Averaged and Denoised Images}
\begin{figure}
    \centering
    \includegraphics[width=\linewidth]{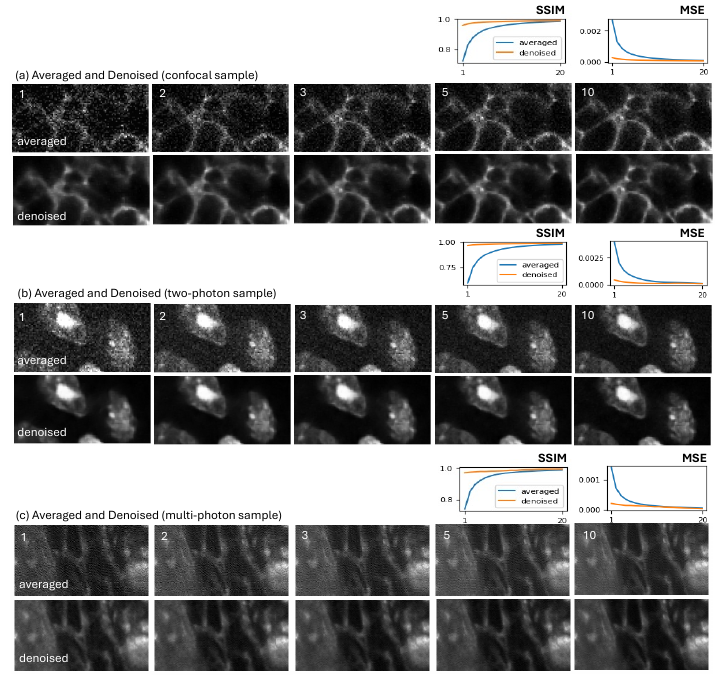}
    \captionsetup{width=1\linewidth}
    \caption{\textbf{Comparison of Averaged vs Denoised Images}: \textbf{a, b, c} demonstrate the different results between averaging and denoising images in confocal, two-photon, and multi-photon samples. Across all three samples, up till 10 noisy images, the denoised predictions consistently outperform the averaged results in both MSE and SSIM. In early iterations, the averaged image is still fairly noisy while from the first denoised prediction, noise is already hard to detect.} 
    \label{supp:avg_denoised}
\end{figure}

How does multi-image denoising compare to simply averaging the noisy measurements? We compare 1-20 image denoising with averaging in Fig.~\ref{supp:avg_denoised} for three different samples. Denoising has a significantly better SSIM and MSE than averaging for few-image denoising across confocal, two-photon, and multi-photon samples. Once 10 or more measurements are used, denoising and averaging become more comparable. This shows that denoising has significant benefits when there are only a few measurements (1-10).  

\section{Model Uncertainty Trends}
\begin{figure}[h!]
    \centering
    \includegraphics[width=\linewidth]{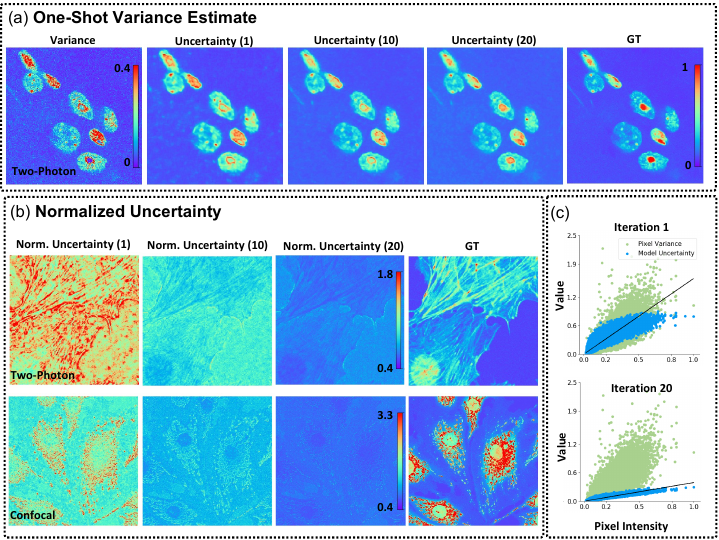}
    \captionsetup{width=1\linewidth}
    \caption{\textbf{Uncertainty is a one-shot estimate of variance (a)}: The uncertainty output of our model with one noisy image closely resembles the structures shown in the variance of twenty noisy images. However, as the number of noisy images to the model increases, the uncertainty output of the model begins to resemble the pixel values of the ground truth image. \textbf{Normalized Uncertainty Converges to the Ground Truth (b)}: In these images, we normalized the values of the uncertainty over the ground truth values. Comparing the normalized uncertainty trends, the uncertainty slowly converges to the ground truth. This is also demonstrated in the scatterplot, where the model uncertainty is shown to categorically decrease up to the last iteration.}
    \label{fig:Uncertainty_relationship}
\end{figure}

What does model uncertainty actually mean? We investigated model uncertainty in relation to image variance and the ground truth image. We found that the uncertainty output of a one-shot denoised image closely resembles the variance of the sample. However, as the noisy images to the denoising model increases, the uncertainty output begins to resemble the ground truth image (Fig.~\ref{fig:Uncertainty_relationship}A). We also find that if we normalize the uncertainty against the ground truth image pixels, the uncertainty will converge to the ground truth as an increasing number of noisy images are passed into the model (Fig.~\ref{fig:Uncertainty_relationship}B). 

To further study the difference between signal intensity and uncertainty, we perform adaptive sampling based on signal intensity and compare this against adaptive sampling based on uncertainty.  Fig.~\ref{fig:Uncertainty_intensity} shows that over 20 iterations, uncertainty-based thresholding reduces the number of rescanned pixels compared to intensity-based thresholding. Over iterations, uncertainty-based thresholding reduces the number of rescanned pixels compared to intensity-based thresholding, potentially saving time and reducing light dose. Across several adaptive interactions, the adaptive masks vary significantly between uncertainty and intensity-based thresholding, showing that uncertainty decreases while intensity remains the same. 

\begin{figure}[h!]
    \centering
    \includegraphics[width=\linewidth]{supp_figures/fig_intensity_vs_uncertainty.pdf}
    \captionsetup{width=1\linewidth}
    \caption{\textbf{Comparison of uncertainty- and intensity-based thresholding during adaptive rescanning}: Over 20 iterations, uncertainty-based thresholding reduces the number of rescanned pixels compared to intensity-based thresholding while achieving comparable image quality and greater savings in light-dose and power. }
    \label{fig:Uncertainty_intensity}
\end{figure}

\section{Multiphoton microscopy dataset}
For our custom multiphoton dataset, we collected measurements from a custom-built inverted scanning microscope~\cite{liu2024deep}. The microscope utilizes a pair of galvanometer mirrors (ScannerMAX Saturn-5 Galvo and Saturn-9 Galvo) to scan the beam, which was focused by a water immersion objective (Olympus XLPLN25XWMP2, 1.05 NA).
A custom-built multimode fiber source at 1100 nm~\cite{liu2024deep} excited THG, SHG, 2PAF, and 3PAF signals, which were subsequently separated by dichroic mirrors (Chroma, T412lpxt, T505lpxr, T570lpxr) and bandpass filters (Chroma, ZET365/20x; Edmund Optics, 84-095; Edmund Optics, 65-159; Semrock, FF01-609/57-25). Photons were collected by four individual photomultiplier tubes (Hamamatsu, H16201), and signals were translated into images using custom-written software. We imaged fixed mouse whisker pad tissue, which was stored in 4\% Paraformaldehyde (PFA) solution at 4 $^\circ$C after excision. Measurements were acquired with a 300-$\mu$m field of view and a pixel dwell time of \SI{1}{\micro\second} for noisy images. Ground truth images were obtained by averaging twenty measurements with a pixel dwell time of \SI{20}{\micro\second}. The delivered pulse energy at the focal plane was approximately 2 nJ. To compile our dataset, we imaged 64 sites within a sample, capturing 20 noisy images and 1 ground truth image per site. Our final dataset contains co-registered 2PAF images from 32 sites for testing and 32 sites for calibration.

\section{Uncertainty prediction without retraining}
\begin{figure}[h!]
    \centering
    \includegraphics[width=\linewidth]{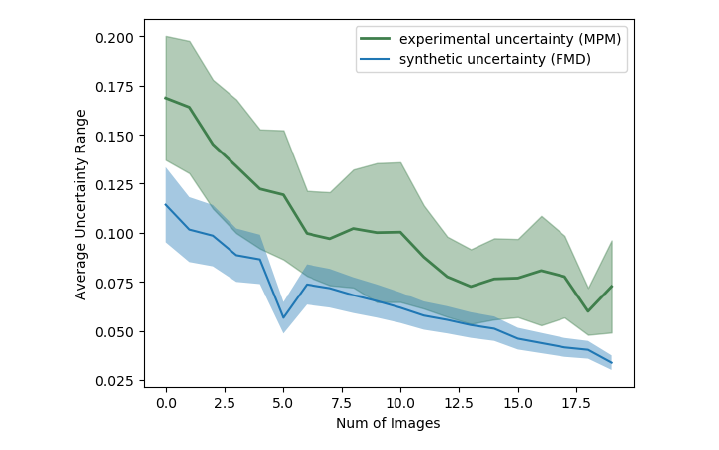}
    \captionsetup{width=1\linewidth}
    \caption{\textbf{Uncertainty interval size without retraining:} We compare the average uncertainty interval size for FMD and MPM measurements using a network trained only on FMD images. While the network is the same, conformal calibration results in a different scaling factor, $\lambda$, between the two modalities. On average, the uncertainty interval is smaller for the FMD measurement than for the MPM measeusent, and this trend is consistent for multi-image denoising with varying numbers of images. Although a network can be used off-the-shelf to perform uncertainty estimation after conformal calibration, the estimated uncertainty estimates may not be optimal. }
    \label{fig:no_model_finetuning}
\end{figure}

One advantage of conformal prediction is that a network trained on one dataset can be used on different data without explicit retraining as long as the conformal calibration step is performed. This means that after conformal calibration using a small dataset of a few images, the statistical guarantees of the uncertainty intervals will hold. However, when a network is not retrained or fine-tuned for a certain imaging modality, the average size of the uncertainty intervals may be larger. In our experiments, we showed that a network trained on the FMD dataset can predict uncertainty sets on experimental MPM measurements after conformal calibration; however, the average uncertainty interval size will be larger. Figure~\ref{fig:no_model_finetuning} illustrates the average uncertainty interval size for the same network using FMD and MPM measurements. The average interval size is larger for MPM since this data has a different distribution. Fine-tuning could enable the network to predict smaller uncertainty intervals, but this may not be practical for every application. For applications with limited data, off-the-shelf networks trained on different data or modalities can be used to predict uncertainty, given the conformal calibration step. For applications where copious data exists, fine-tuning or retraining a network that is specific to the application can result in tighter bounds.

\section{Measurement order}
As we highlighted in Fig.~\ref{supp:hallucin_single}, different noisy measurements of the same sample can result in different predictions and hallucinations. This suggests that different measurements of the same sample could result in slightly different uncertainty estimates and adaptive imaging progressions. To further investigate this, we simulated adaptive imaging using measurements with different randomized orders, Fig.~\ref{fig:ordering}. We investigated whether changing the acquisition order of the measurements would influence the final image estimate and adaptive uncertainty map. For a typical adaptive thresholding scheme in which 30\% of the initial pixels are resampled, the final image predictions appear consistent regardless of the measurement order.  
\begin{figure}[h!]
    \centering
    \includegraphics[width=\linewidth]{supp_figures/fig_reordering.pdf}
    \captionsetup{width=1\linewidth}
    \caption{\textbf{Effect of measurement order on adaptive imaging:} Adaptive sampling predictions and masking patterns using randomized measurement orders, assuming resampling 30\% of the initial pixels. Randomizing the measurements appears to lead to consistent predictions and adaptive masks. Across different randomized measurement orders, final results varied between 1-3\%.}
    \label{fig:ordering}
\end{figure}
\bibliography{bib}